\documentclass[aps,prl,twocolumn,10pt,superscriptaddress]{revtex4-1}

\usepackage{graphicx}
\usepackage[T1]{fontenc}

\begin{document}

\newcommand{\AlOx}{$\alpha$-Al$_{2}$O$_{3}$}
\newcommand{\MFe}{$^{57}$Fe}
\newcommand{\dgr}{^{\circ}}


\title{Disentangling magnetic order on nanostructured surfaces}


\author{D. Erb}
\email[]{d.erb@hzdr.de}
\altaffiliation{present address: Helmholtz-Zentrum Dresden-Rossen\-dorf, Bautzner Landstra{\ss}e 400, 01328 Dresden, Germany}
\affiliation{Deutsches Elektronen-Synchrotron DESY, Notkestra{\ss}e 85, 22607 Hamburg, Germany}

\author{K. Schlage}
\affiliation{Deutsches Elektronen-Synchrotron DESY, Notkestra{\ss}e 85, 22607 Hamburg, Germany}

\author{L. Bocklage}
\affiliation{Deutsches Elektronen-Synchrotron DESY, Notkestra{\ss}e 85, 22607 Hamburg, Germany}
\affiliation{The Hamburg Centre for Ultrafast Imaging, Luruper Chaussee 149, 22761 Hamburg, Germany}

\author{R. H\"{u}bner}
\affiliation{Helmholtz-Zentrum Dresden-Rossendorf, Bautzner Landstra{\ss}e 400, 01328 Dresden, Germany}

\author{D. G. Merkel}
\altaffiliation{on leave from Wigner Research Centre for Physics, Hungarian Academy of Sciences, H-1525 Budapest, Hungary}
\affiliation{European Synchrotron Radiation Facility, 71 avenue des Martyrs, 38000 Grenoble, France}

\author{R. R\"{u}ffer}
\affiliation{European Synchrotron Radiation Facility, 71 avenue des Martyrs, 38000 Grenoble, France}

\author{H.-C. Wille}
\affiliation{Deutsches Elektronen-Synchrotron DESY, Notkestra{\ss}e 85, 22607 Hamburg, Germany}

\author{R. R\"{o}hlsberger}
\affiliation{Deutsches Elektronen-Synchrotron DESY, Notkestra{\ss}e 85, 22607 Hamburg, Germany}
\affiliation{The Hamburg Centre for Ultrafast Imaging, Luruper Chaussee 149, 22761 Hamburg, Germany}


\date{\today}

\begin{abstract}
We present a synchrotron-based X-ray scattering technique which allows disentangling magnetic properties of hetero\-geneous systems with nanopatterned surfaces. This technique combines the nm-range spatial resolution of surface morphology features provided by Grazing Incidence Small Angle X-ray Scattering and the high sensitivity of Nuclear Resonant Scattering to magnetic order. A single experiment thus allows attributing magnetic properties to structural features of the sample; chemical and structural properties may be correlated analogously. We demonstrate how this technique shows the correlation between structural growth and evolution of magnetic properties for the case of a remarkable magnetization reversal in a structurally and magnetically nanopatterned sample system.
\end{abstract}


\maketitle


Knowledge of the relations between morphology and physical properties of nano-scaled objects is key to engineer the functionalities of devices in nanotechnology. Prominent examples are the size- and shape-dependent magnetic properties of nanoparticles \cite{Mehdaoui11, Reddy12}, which are highly relevant for medical diagnostics and therapy, of ferromagnetic components in magnonic devices \cite{Topp10} and magnetoplasmonic systems \cite{Armelles13}, or the size- and composition-dependent reactivity and morphological changes of catalytically active nanoparticles during chemical reaction \cite{Wettergren14, Hejral16}. Thus, methods yielding the nanostructure morphology with high spatial resolution and enabling comprehensive chemical characterization or delivering precise information on magnetic properties are indispensable tools in nanoscience: Physicochemical characterization is routinely accomplished by methods such as high-resolution transmission electron microscopy \cite{Li12}, grazing incidence small angle X-ray scattering \cite{Rauscher99}, mass spectroscopy \cite{Park05}, dynamic light scattering \cite{Pecora00}, surface plasmon resonance \cite{Juve13}, or spectroscopic methods using radiation from infrared to X-ray wavelengths \cite{Li12, Rauscher99, Park05, Pecora00, Juve13, Shukla03, Amendola09, Prieto12}. For magnetic nanostructure characterization a variety of techniques has been established, all with unique assets but also with certain drawbacks. Scanning probe techniques provide very high spatial resolution \cite{Wiesendanger09}, but are insensitive to magnetization dynamics. Kerr and Faraday microscopy offer picosecond time resolution, but their spatial resolution is merely in the sub-micrometer range \cite{McCord15}. Scanning electron microscopy with polarization analysis measures the magnetization vector orientation directly via the spin polarization of secondary electrons \cite{Koike13} and x-ray photoemission electron microscopy combines good spatial and temporal resolution with element specificity \cite{Cheng12}. Being based on the detection of secondary electrons, however, both require ultra high vacuum conditions and only allow for applying very weak or localized magnetic fields to the sample. Diffraction magneto-optical Kerr effect measurements are simple to realize and can accommodate various sample environments \cite{Grimsditch04}, but the wavelength of the employed light makes nanostructure characterization unfeasible. Neutron scattering techniques can provide both structural and magnetic information \cite{Zabel07}, but suffer from the low neutron fluxes, which limits the possibilities for in-situ measurements, e.g. during nanostructure growth. Methods based on x-ray transmission or scattering in transmission geometry (scanning transmission x-ray microscopy, transmission imaging x-ray microscopy, or x-ray holography) offer element-specificity and high spatial resolution, but pose constraints on sample environment volumes, require a transmissive substrate, or require the sample to be processed by microfabrication techniques \cite{Stoehr93, Turner11}.

We propose a method for investigating nano-scaled objects, which can simultaneously deliver morphological parameters with sub-nanometer lateral precision and the static and dynamic magnetic characteristics of these structures under in-situ conditions of growth, high temperatures, reactive environments, or strong magnetic fields. To achieve this, we combine two X-ray scattering techniques into a single experiment, namely Grazing Incidence Small Angle X-ray Scattering (GISAXS) and Nuclear Resonant Scattering (NRS). GISAXS provides morphological characterization of nanometer-sized surface features, based on the angular distribution of scattered photons depending on the sample structure. Thus, three-dimensional shapes and lateral arrangements of nanostructures supported on surfaces or buried in thin films are obtained with nanometer resolution. NRS yields information on magnetic ordering, enables precise determination of in-plane and out-of-plane magnetic moment orientation and allows for sensitive detection of magnetization dynamics with an accuracy of a few degrees and sub-microsecond time resolution \cite{Roehlsberger99, Roehlsberger03, Schlage13, Bocklage15} by probing the coherent elastic resonant scattering of photons from M\"{o}ssbauer-active nuclei \cite{Roehlsberger04}. Intensity maxima in a GISAXS pattern originate from photons, which are scattered off different periodically repeated structural components of the sample. Photons, which have been resonantly scattered from nuclei, are identified by their time delay with respect to the photons, which have been non-resonantly scattered from electrons. The coherent elastic nuclear resonant scattering of photons results in a characteristic time spectrum of the detected intensity. At the specular intensity maximum, this time spectrum reflects the integrated magnetic properties of the entire sample \cite{Schlage12, Sharma15}. By placing the detector at selected off-specular intensity maxima within the pattern, however, one obtains information on the magnetic properties of the specific structural component of the sample which the selected intensity maximum is related to. Thus, both structural information and site-specific magnetic characteristics are gathered simultaneously, directly revealing the correlations between these properties. 

From merging GISAXS and NRS into a single technique referred to as Grazing Incidence Nuclear Small Angle X-ray Scattering (GINSAXS), comprehensive structural and magnetic information on heterogeneous systems with periodic nanoscopic surface morphology can be obtained. In this paper we provide an experimental proof of principle for the aforementioned concept under two different in-situ conditions. While we apply GINSAXS to disentangle heterogeneous magnetic properties, the method could also be employed to elucidate heterogeneous chemical composition or crystal structure: The hyperfine interactions probed by NRS also characterize local chemical environments and local lattice structures of resonant nuclei \cite{Vogl09, Potapkin13, Freindl13}. Generally, a combination of GISAXS with NRS is applicable to samples with non-planar surface or interface morphologies, such as periodic arrangements of uniform nanostructures with tilted or curved surfaces. Such a technique could be highly beneficial for studying facet-selective adsorption \cite{Chiu13, Mdluli11}, overgrowth \cite{Sun12, Liu11, Mankin15}, or reactivity \cite{Harn15, Li13}. It could also help to clarify the development of magnetic order and the magnetization reversal in faceted nanoparticles with magnetic shells \cite{Nasirpouri11}, or serve to study the magnetic properties of nanoparticle-based mesocrystals \cite{Disch11, Chen13, Josten17}.

\begin{figure}
\includegraphics[width=\linewidth, trim={0cm, 0cm, 0cm, 0cm}, clip]{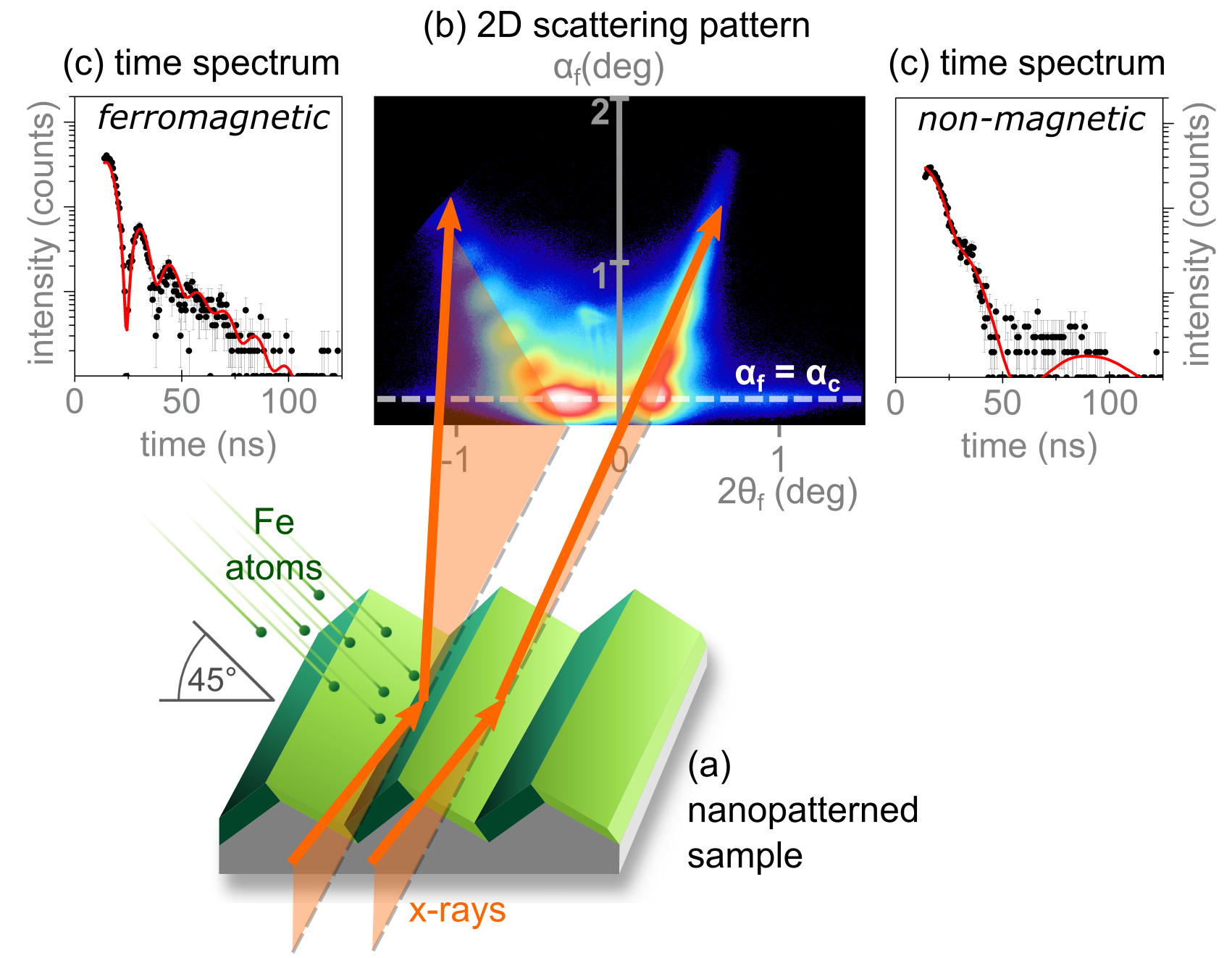}%
\caption{\footnotesize{Illustration of the GINSAXS principle by means of a sample with faceted surface: (a) A heterogeneous Fe film is grown on a nanofaceted substrate by sputter deposition under non-normal incidence. (b) The 2D scattering pattern is the same for both non-resonantly and resonantly scattered photons; photons detected in the left (right) crystal truncation rod (CTR) carry information on the Fe film regions on the R-plane (S-plane) facets. (c) Using a time-resolving detector to record only resonant photons, nuclear resonant time spectra are taken at the left and right CTR, evidencing the different magnetic properties of the Fe film on the R-plane and S-plane facets, respectively.} 
\label{fig:Figure1}}
\end{figure}

We demonstrate the capability of GINSAXS by investigating a nanostructured sample system with periodically varying morphological and magnetic properties under in-situ conditions requiring ultra-high vacuum and external magnetic fields, respectively. The sample consists of an \AlOx~substrate with parallel nanometer-scale facets (see Supplemental Material \cite{Supplement} and \cite{Heffelfinger97}), supporting a thin continuous Fe film. The substrate has an average facet height of $h = 15$~nm and period of $L = 80$~nm, the average facet tilt angles are $\beta_{R} = 30\dgr$ and $\beta_{S} = 17\dgr$. From these values, average facet widths of $w_{R} \approx 32$~nm and $w_{S} \approx 55$~nm are calculated. The Fe film is grown by room temperature sputter deposition from a polar angle of $45\dgr$ and an azimuthal angle of $90\dgr$ with respect to the facet edges. Here, the R-plane facets are facing the sputtering source, so that the deposition rate is higher on these facets than on the S-plane facet, which are avert from the source. Consequently, the Fe film consists of thicker regions (18~nm) on the narrower R-plane facets and thinner regions (13~nm) on the wider S-plane facets (see Fig.~\ref{fig:Figure1}(a)). As determined from a GISAXS pattern recorded with the facet edges aligned perpendicular to the incident X-ray beam, the Fe film is polycrystalline with a crystallite size of approximately 5~nm (see Supplemental Material \cite{Supplement}). To prevent oxidation of the Fe film, the sample was capped with a Cr layer. The corrugated shape of the Fe film induces a uniaxial magnetic anisotropy with the easy axis of magnetization parallel to the substrate facet edges \cite{Liedke13}.

As summarized in Fig.~\ref{fig:Figure1}, GINSAXS is conducted in the following sequence: First, a conventional 2D GISAXS pattern is recorded at a suitable angle of incidence using an area detector. With the facet edges aligned parallel to the incident X-ray beam, the GISAXS pattern is characterized by two crystal truncation rods (CTRs) originating from the R-plane and S-plane facets, respectively \cite{Rauscher99}. Second, the positions of intensity maxima which are specific for certain structural units of the sample are selected from the GISAXS pattern. Here, these are the positions of highest intensity along the two CTRs. The angle of incidence is adjusted to maximize the resonantly scattered intensity. Third, using a time-resolving point detector (avalanche photo diode, APD) nuclear resonant time spectra are recorded at the selected positions. 
The angular distribution of scattered intensity measured in the GISAXS pattern is a signature of the nanometer-scale surface morphology of the sample: the two tilted scattering rods correspond to the Fe film regions supported by the substrate facets with R-plane and S-plane orientation, respectively. Intensity modulations along the CTRs are related to the Fe film thicknesses on the respective facet surfaces (similar to Kiessig fringes). The film thicknesses and geometrical parameters describing the sample morphology are obtained by simulating the GISAXS patterns using the program FitGISAXS \cite{Babonneau10}. The nuclear resonant time spectra serve as fingerprints of the magnetic characteristics of the different repeat units of the sample, conveying information on the degree of magnetic order and the magnetization orientation. The shape of a time spectrum correlates with these properties via strength and orientation, respectively, of the magnetic hyperfine field $\textbf{B}_{hf}$ at the Fe nuclei. Time spectra were fitted using the program CONUSS \cite{Sturhahn00} (see Supplemental Material \cite{Supplement}).

Following the procedure described above, GINSAXS was performed at an X-ray energy of 14.4~keV, i.e. the resonance energy of \MFe, at the high resolution dynamics beamline P01 at \mbox{PETRA III} and the nuclear resonance beamline ID18 at ESRF (see Supplemental Material \cite{Supplement}). We performed  two independent in-situ GINSAXS experiments on a sample system which is structurally and magnetically heterogeneous on the nanoscale. The experiments provide spatially resolved information on the magnetization reversal and allow correlating film growth and development of ferromagnetic ordering.

\begin{figure}
\includegraphics[width=0.95\linewidth, trim={0cm, 0cm, 0cm, 0cm}, clip]{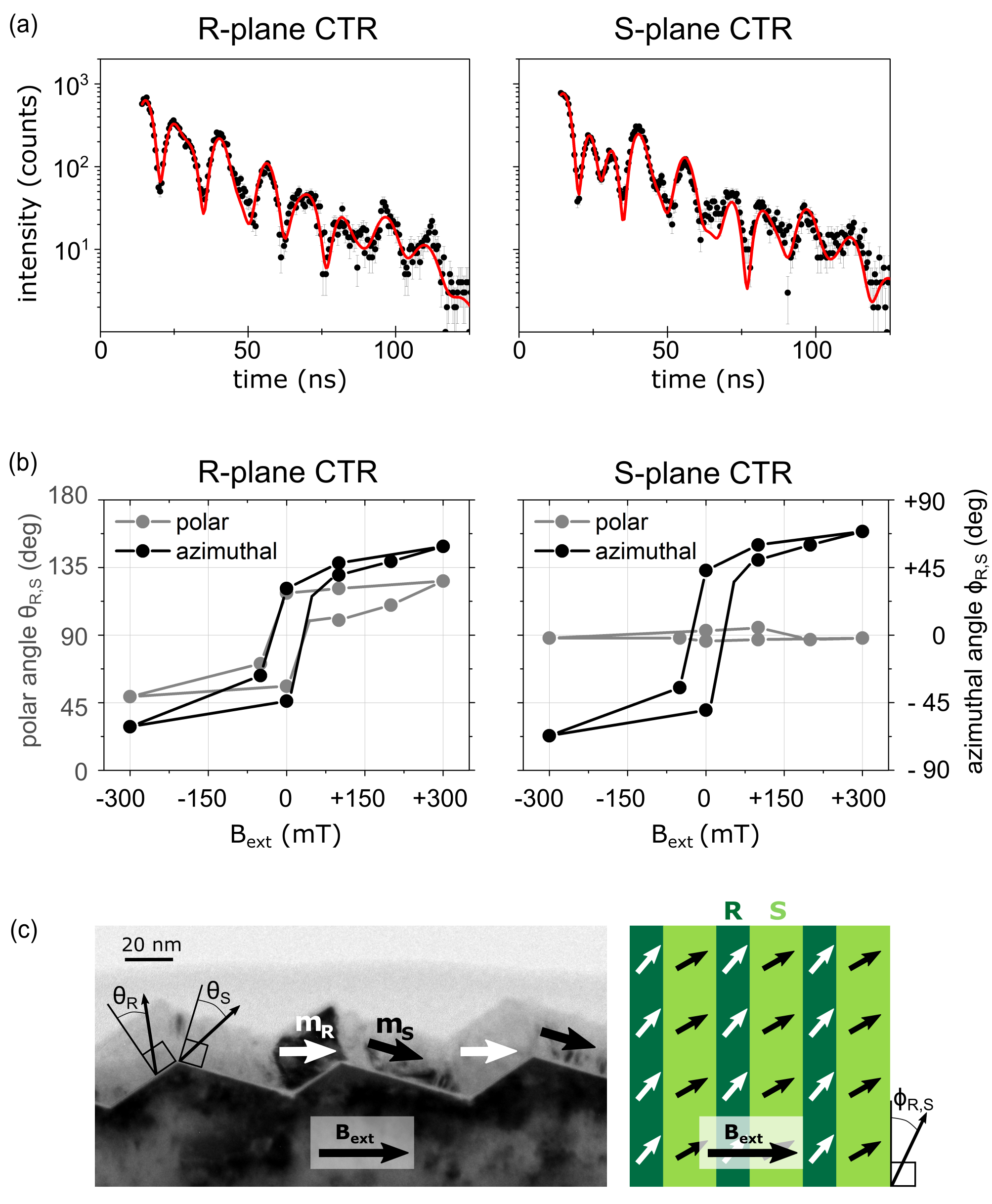}%
\caption{\footnotesize{(a) Exemplary NRS time spectra and corresponding fits (red) for the Fe film regions on R- and S-plane facets, respectively, in an applied magnetic field of $B_{ext} = 200$~mT. The differences in the beat patterns evidence unequal magnetization orientations. (b) Azimuthal (in-plane) and polar (out-of-plane) magnetization orientation in R-plane and S-plane film regions as extracted from fits of the NRS time spectra. Solid lines are guides to the eye. (c) A cross-sectional transmission electron micrograph (left) and a top-view sketch (right) of the sample, with arrows indicating the out-of-plane and in-plane components of the magnetization vectors relative to external magnetic field.} 
\label{fig:Figure2}}
\end{figure}

In the first experiment we resolve the heterogeneous magnetization reversal in the nanostructured Fe film upon applying an external magnetic field. The polar and azimuthal angles $\theta_{R,S}$ and $\phi_{R,S}$ of the magnetization as extracted from fitting the NRS time spectra for both R-plane and S-plane film regions are plotted as functions of the external magnetic field in Fig.~\ref{fig:Figure2}(b). The angles are defined independently for R-plane and S-plane film regions, such that the magnetization is in the plane of the respective film region for a polar angle of $\theta_{R,S} = 90\dgr$ and parallel to the facet edges for an azimuthal angle of $\phi_{R,S} = 0\dgr$. The hysteretic behavior of the azimuthal magnetization orientation is similar for the R-plane and S-plane regions of the Fe film: the magnetization is displaced from the easy axis parallel to the facet edges to similar extents but does not align fully with the orientation of the external magnetic field. This can be accounted for by the pronounced uniaxial in-plane magnetic anisotropy of the uniaxially corrugated film.
The R-plane and S-plane regions of the Fe film differ markedly, however, in the polar  magnetization orientation \cite{Liedke13}: In the S-plane Fe film regions the polar magnetization orientation remains almost constant at $\theta_{S} = 90\dgr$, i.e. it remains parallel to the film plane of these regions even at highest field strength. In contrast, the magnetization in the R-plane regions is deflected to an orientation in between the direction of the external magnetic field at $\theta = 60\dgr$ and $\theta = 120\dgr$, respectively, and the magnetization in the S-plane film regions at $\theta = 43\dgr$ and $\theta = 137\dgr$, respectively. The film regions on the S-plane facets are both thinner and wider than those on the R-plane facets. Thus, the shape anisotropy is more pronounced and in-plane orientation of the magnetization is preferred in the S-plane film regions. Furthermore, the external magnetic field is applied parallel to the average sample surface and does not enclose the same angle with the R-plane and S-plane facets. Consequently, the magnitude of the external magnetic field component normal to the film plane is by approximately 70\% larger for the R-plane film regions on the R-plane facets. Interface coupling between the Fe film and the antiferromagnetic Cr capping layer with high magnetic anisotropy may be a cause for the inertness observed in the magnetization returning to its easy axis orientation at remanence. The influence of this effect on the measurement is strong due to the surface sensitivity of NRS at $\alpha_{i} = 0.16^{\circ}$. NRS spectra taken at higher incidence angles indicate a spring-like magnetization structure of the Fe film, where the top layers of the Fe film are coupled to the Cr capping layer, while the bottom layers are free to relax toward the easy axis orientation when no external field is applied (see Supplemental Material \cite{Supplement}).

\begin{figure}
\includegraphics[width=0.95\linewidth, trim={0.25cm, 0cm, 0cm, 0cm}, clip]{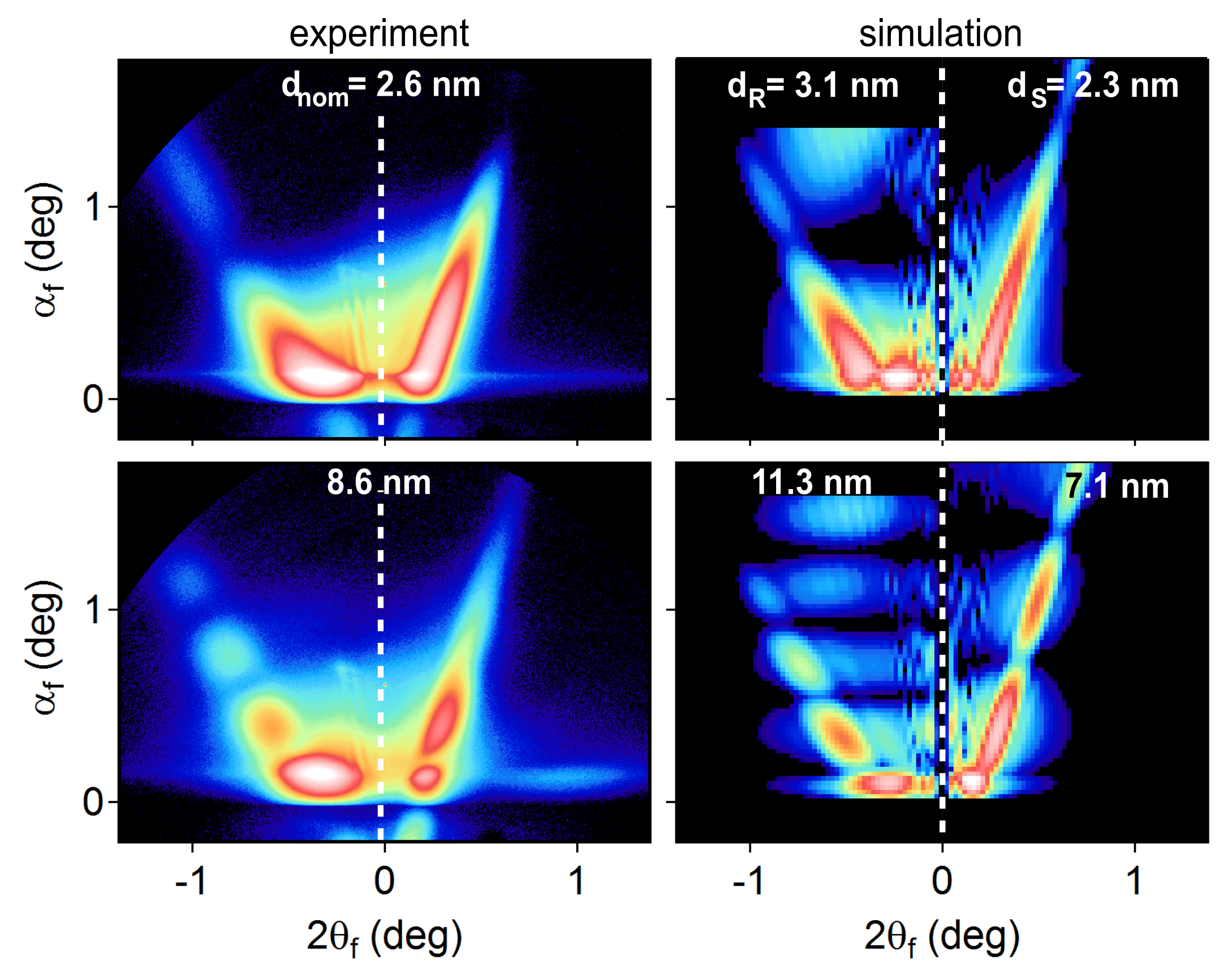}%
\caption{\footnotesize{Exemplary non-resonant GISAXS patterns and corresponding simulations for two stages of Fe deposition onto the faceted substrate. The dashed lines indicate the specular scattering plane. The GISAXS patterns are dominated by the tilted crystal truncation rods (CTRs) originating from the surface facets. The different periods of the intensity modulations along the CTRs result from the different thicknesses of the Fe film on the R-plane and S-plane facets, respectively. Labels state the nominally deposited Fe film thickness $\mathrm{d_{nom}}$ and the thicknesses of the film regions $\mathrm{d_{R}}$ and $\mathrm{d_{S}}$ as obtained from simulations.} 
\label{fig:Figure3}}
\end{figure}

\begin{figure}
\includegraphics[width=0.95\linewidth, trim={0.25cm, 0cm, 0.5cm, 0cm}, clip]{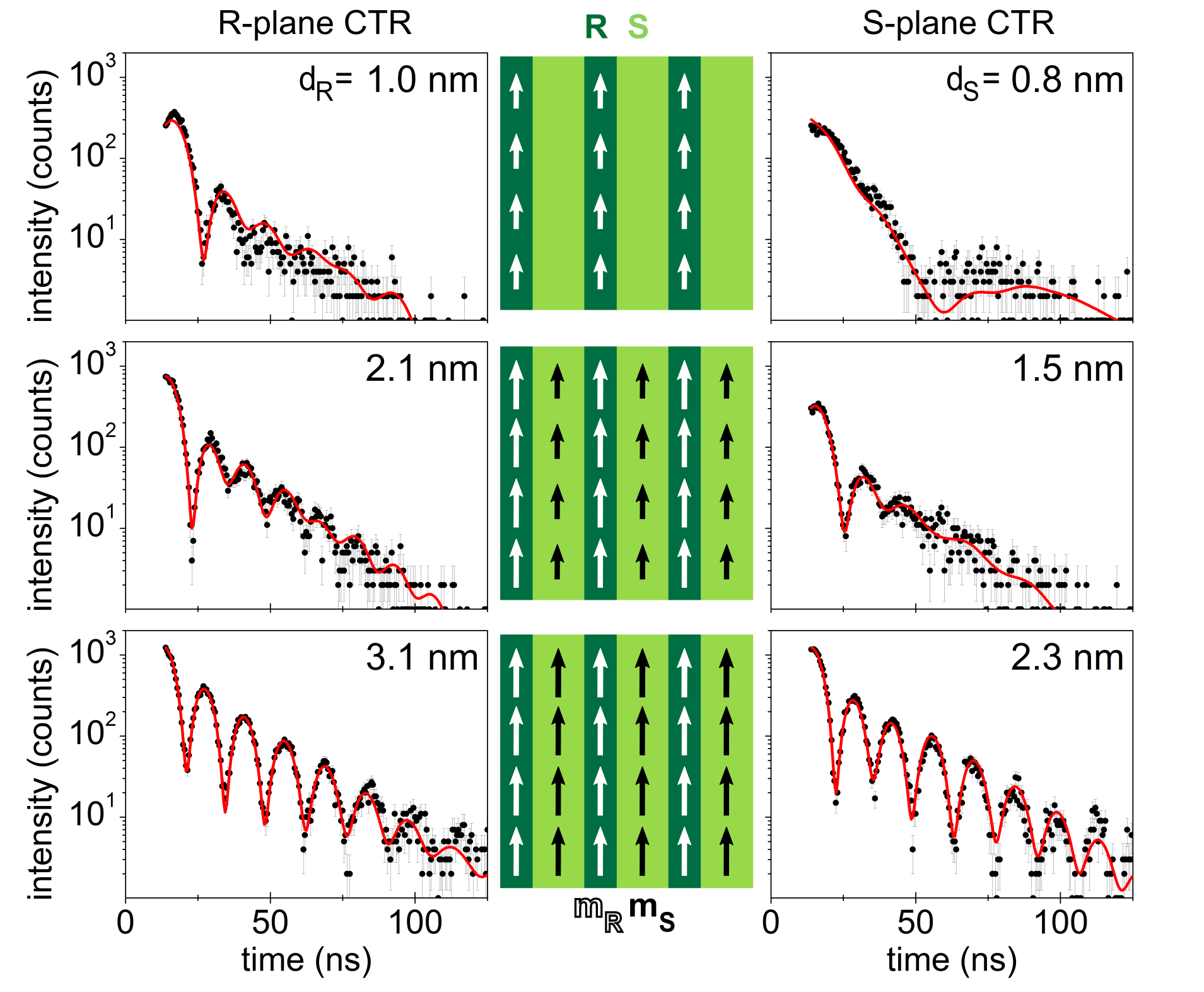}%
\caption{\footnotesize{Sequences of NRS time spectra (top to bottom) with corresponding fits (red) for film regions on R-plane and S-plane facets at subsequent stages of Fe deposition. The development of the characteristic beat pattern evidences the successive transition from a non-magnetic to a ferromagnetic state in the R-plane and S-plane regions. Strength and orientation of magnetic moments in the film regions on R-plane and S-plane facets are indicated by black and white arrows, respectively.} 
\label{fig:Figure4}}
\end{figure}

Furthermore, we observed the correlation of growth and magnetic stabilization in the stripe-like regions of the Fe film in an in-situ experiment during Fe deposition. Exemplary GISAXS patterns taken during growth of the Fe film are compared to the corresponding simulations in Fig.~\ref{fig:Figure3} (see Supplemental Material \cite{Supplement}). The periods of intensity modulation along the CTRs decrease in correspondence to the increasing film thicknesses; the unequal modulation periods for each deposition stage evidence the different thicknesses of the Fe film on S-plane and R-plane facets. 
Simultaneously to the film growth observed in the GISAXS patterns, the evolution of magnetic properties is evidenced by characteristic changes in the time spectra. Fig.~\ref{fig:Figure4} depicts sequences of time spectra for the film regions on R-plane and S-plane facets recorded during growth and sketches the corresponding strength and orientation of magnetic moments. A non-magnetic state is characterized by the lack of a beat pattern on the time spectrum, while the time spectrum of a ferromagnetic state exhibits a pronounced beat pattern. The sequences show the successive evolution of beat patterns in the time spectra for the different film regions, thus evidencing a consecutive transition from a non-magnetic to a ferromagnetic state first in the R-plane film regions, then in S-plane regions. Notably, a state is observed in which the S-plane film regions are still non-magnetic, while the R-plane regions already exhibit a magnetization parallel to the facet edges. Full ferromagnetic order is observed at a film thickness of $(2.8 \pm 0.1)$~nm for the R-plane Fe film regions. In the S-plane regions, ferro\-magnetic order is established at a thickness of $(2.3 \pm 0.1)$~nm already, which may be due to the direct contact with the fully ferromagnetic R-plane film regions. In polycrystalline Fe thin films with uniaxially corrugated shape, ferromagnetic ordering was observed at film thicknesses 1~nm~$<$ d $<$~3~nm; the results of this experiment thus agree well with former findings \cite{Shiratsuchi04}. For thicknesses larger than 2.8~nm and 2.3~nm, respectively, all film regions on the R-plane facets have a magnetic hyperfine field magnitude close to the $\alpha$-Fe bulk value. Both the R-plane and S-plane film regions now show time spectra with pronounced beat patterns evidencing ferromagnetically ordered Fe with the magnetization oriented exactly parallel to the direction of the incoming beam, i.e. parallel to the facet edges. Cross-sectional transmission electron microscopy (see Fig.\ref{fig:Figure2}(c)) confirms that the free surface of the Fe film and its interface with the substrate are parallel as indicated by the sharp intensity modulations in the GISAXS patterns.

In conclusion, GINSAXS, a combination of GISAXS and NRS, allows detecting the magnetic information from different structural units of a nanostructured sample separately. This approach has a great potential for characterizing corrugated magnetic materials, which are of interest for applications in magnetic sensing due to their shape-induced uniaxial magnetic anisotropy. Also standing spin waves in magnetic lattices may be investigated. Furthermore, it is conceivable to employ it for studying supported or buried nanoparticles with faceted or curved surfaces or with core-shell morphologies. The successful demonstration of GINSAXS may also encourage testing the feasibility of combining X-ray Absorption Near Edge Structure (XANES) and GISAXS in an analogous manner: This would be a powerful tool for investigating nanostructures with shape-dependent heterogeneous chemical properties, for instance for in-situ studies of catalytic processes with facet-selective reactivity.

\begin{acknowledgments}
We gratefully acknowledge David Babonneau, Universit\'{e} de Poitiers, for implementing a new form factor in FitGISAXS on our request, which enabled simulating the GISAXS patterns of the Fe film on the faceted substrate.
We thank Shengqiang Zhou, Helmholtz-Zentrum Dresden-Rossendorf, for measuring VSM hysteresis curves of the sample.
\end{acknowledgments}

%


\begin{thebibliography}{49}%
\makeatletter
\providecommand \@ifxundefined [1]{%
 \@ifx{#1\undefined}
}%
\providecommand \@ifnum [1]{%
 \ifnum #1\expandafter \@firstoftwo
 \else \expandafter \@secondoftwo
 \fi
}%
\providecommand \@ifx [1]{%
 \ifx #1\expandafter \@firstoftwo
 \else \expandafter \@secondoftwo
 \fi
}%
\providecommand \natexlab [1]{#1}%
\providecommand \enquote  [1]{``#1''}%
\providecommand \bibnamefont  [1]{#1}%
\providecommand \bibfnamefont [1]{#1}%
\providecommand \citenamefont [1]{#1}%
\providecommand \href@noop [0]{\@secondoftwo}%
\providecommand \href [0]{\begingroup \@sanitize@url \@href}%
\providecommand \@href[1]{\@@startlink{#1}\@@href}%
\providecommand \@@href[1]{\endgroup#1\@@endlink}%
\providecommand \@sanitize@url [0]{\catcode `\\12\catcode `\$12\catcode
  `\&12\catcode `\#12\catcode `\^12\catcode `\_12\catcode `\%12\relax}%
\providecommand \@@startlink[1]{}%
\providecommand \@@endlink[0]{}%
\providecommand \url  [0]{\begingroup\@sanitize@url \@url }%
\providecommand \@url [1]{\endgroup\@href {#1}{\urlprefix }}%
\providecommand \urlprefix  [0]{URL }%
\providecommand \Eprint [0]{\href }%
\providecommand \doibase [0]{http://dx.doi.org/}%
\providecommand \selectlanguage [0]{\@gobble}%
\providecommand \bibinfo  [0]{\@secondoftwo}%
\providecommand \bibfield  [0]{\@secondoftwo}%
\providecommand \translation [1]{[#1]}%
\providecommand \BibitemOpen [0]{}%
\providecommand \bibitemStop [0]{}%
\providecommand \bibitemNoStop [0]{.\EOS\space}%
\providecommand \EOS [0]{\spacefactor3000\relax}%
\providecommand \BibitemShut  [1]{\csname bibitem#1\endcsname}%
\let\auto@bib@innerbib\@empty
\bibitem [{\citenamefont {Mehdaoui}\ \emph {et~al.}(2011)\citenamefont
  {Mehdaoui}, \citenamefont {Meffre}, \citenamefont {Carrey}, \citenamefont
  {Lachaize}, \citenamefont {Lacroix}, \citenamefont {Gougeon}, \citenamefont
  {Chaudret},\ and\ \citenamefont {Respaud}}]{Mehdaoui11}%
  \BibitemOpen
  \bibfield  {author} {\bibinfo {author} {\bibfnamefont {B.}~\bibnamefont
  {Mehdaoui}}, \bibinfo {author} {\bibfnamefont {A.}~\bibnamefont {Meffre}},
  \bibinfo {author} {\bibfnamefont {J.}~\bibnamefont {Carrey}}, \bibinfo
  {author} {\bibfnamefont {S.}~\bibnamefont {Lachaize}}, \bibinfo {author}
  {\bibfnamefont {L.-M.}\ \bibnamefont {Lacroix}}, \bibinfo {author}
  {\bibfnamefont {M.}~\bibnamefont {Gougeon}}, \bibinfo {author} {\bibfnamefont
  {B.}~\bibnamefont {Chaudret}}, \ and\ \bibinfo {author} {\bibfnamefont
  {M.}~\bibnamefont {Respaud}},\ }\href@noop {} {\bibfield  {journal} {\bibinfo
   {journal} {{Adv. Funct. Mater.}}\ }\textbf {\bibinfo {volume} {21}},\
  \bibinfo {pages} {4573} (\bibinfo {year} {2011})}\BibitemShut {NoStop}%
\bibitem [{\citenamefont {Reddy}\ \emph {et~al.}(2012)\citenamefont {Reddy},
  \citenamefont {Arias}, \citenamefont {Nicolas},\ and\ \citenamefont
  {Couvreur}}]{Reddy12}%
  \BibitemOpen
  \bibfield  {author} {\bibinfo {author} {\bibfnamefont {L.~H.}\ \bibnamefont
  {Reddy}}, \bibinfo {author} {\bibfnamefont {J.~L.}\ \bibnamefont {Arias}},
  \bibinfo {author} {\bibfnamefont {J.}~\bibnamefont {Nicolas}}, \ and\
  \bibinfo {author} {\bibfnamefont {P.}~\bibnamefont {Couvreur}},\ }\href@noop
  {} {\bibfield  {journal} {\bibinfo  {journal} {Chem. Rev.}\ }\textbf
  {\bibinfo {volume} {112}},\ \bibinfo {pages} {5818} (\bibinfo {year}
  {2012})}\BibitemShut {NoStop}%
\bibitem [{\citenamefont {Topp}\ \emph {et~al.}(2010)\citenamefont {Topp},
  \citenamefont {Heitmann}, \citenamefont {Kostylev},\ and\ \citenamefont
  {Grundler}}]{Topp10}%
  \BibitemOpen
  \bibfield  {author} {\bibinfo {author} {\bibfnamefont {J.}~\bibnamefont
  {Topp}}, \bibinfo {author} {\bibfnamefont {D.}~\bibnamefont {Heitmann}},
  \bibinfo {author} {\bibfnamefont {M.~P.}\ \bibnamefont {Kostylev}}, \ and\
  \bibinfo {author} {\bibfnamefont {D.}~\bibnamefont {Grundler}},\ }\href@noop
  {} {\bibfield  {journal} {\bibinfo  {journal} {Phys. Rev. Lett.}\ }\textbf
  {\bibinfo {volume} {104}},\ \bibinfo {pages} {207205} (\bibinfo {year}
  {2010})}\BibitemShut {NoStop}%
\bibitem [{\citenamefont {Armelles}\ \emph {et~al.}(2013)\citenamefont
  {Armelles}, \citenamefont {Cebollada}, \citenamefont
  {Garc\'{i}a-Mart\'{i}n},\ and\ \citenamefont {Gonz\'{a}lez}}]{Armelles13}%
  \BibitemOpen
  \bibfield  {author} {\bibinfo {author} {\bibfnamefont {G.}~\bibnamefont
  {Armelles}}, \bibinfo {author} {\bibfnamefont {A.}~\bibnamefont {Cebollada}},
  \bibinfo {author} {\bibfnamefont {A.}~\bibnamefont {Garc\'{i}a-Mart\'{i}n}},
  \ and\ \bibinfo {author} {\bibfnamefont {M.~U.}\ \bibnamefont
  {Gonz\'{a}lez}},\ }\href@noop {} {\bibfield  {journal} {\bibinfo  {journal}
  {Adv. Optical Mater.}\ }\textbf {\bibinfo {volume} {1}},\ \bibinfo {pages}
  {10} (\bibinfo {year} {2013})}\BibitemShut {NoStop}%
\bibitem [{\citenamefont {Wettergren}\ \emph {et~al.}(2014)\citenamefont
  {Wettergren}, \citenamefont {Schweinberger}, \citenamefont {Deiana},
  \citenamefont {Ridge}, \citenamefont {Crampton}, \citenamefont {R\"{o}tzer},
  \citenamefont {Hansen}, \citenamefont {Zhdanov}, \citenamefont {Heiz},\ and\
  \citenamefont {Langhammer}}]{Wettergren14}%
  \BibitemOpen
  \bibfield  {author} {\bibinfo {author} {\bibfnamefont {K.}~\bibnamefont
  {Wettergren}}, \bibinfo {author} {\bibfnamefont {F.~F.}\ \bibnamefont
  {Schweinberger}}, \bibinfo {author} {\bibfnamefont {D.}~\bibnamefont
  {Deiana}}, \bibinfo {author} {\bibfnamefont {C.~J.}\ \bibnamefont {Ridge}},
  \bibinfo {author} {\bibfnamefont {A.~S.}\ \bibnamefont {Crampton}}, \bibinfo
  {author} {\bibfnamefont {M.~D.}\ \bibnamefont {R\"{o}tzer}}, \bibinfo
  {author} {\bibfnamefont {T.~W.}\ \bibnamefont {Hansen}}, \bibinfo {author}
  {\bibfnamefont {V.~P.}\ \bibnamefont {Zhdanov}}, \bibinfo {author}
  {\bibfnamefont {U.}~\bibnamefont {Heiz}}, \ and\ \bibinfo {author}
  {\bibfnamefont {C.}~\bibnamefont {Langhammer}},\ }\href@noop {} {\bibfield
  {journal} {\bibinfo  {journal} {Nano Lett.}\ }\textbf {\bibinfo {volume}
  {14}},\ \bibinfo {pages} {5803} (\bibinfo {year} {2014})}\BibitemShut
  {NoStop}%
\bibitem [{\citenamefont {Hejral}\ \emph {et~al.}(2016)\citenamefont {Hejral},
  \citenamefont {M\"{u}ller}, \citenamefont {Balmes}, \citenamefont {Pontoni},\
  and\ \citenamefont {Stierle}}]{Hejral16}%
  \BibitemOpen
  \bibfield  {author} {\bibinfo {author} {\bibfnamefont {U.}~\bibnamefont
  {Hejral}}, \bibinfo {author} {\bibfnamefont {P.}~\bibnamefont {M\"{u}ller}},
  \bibinfo {author} {\bibfnamefont {O.}~\bibnamefont {Balmes}}, \bibinfo
  {author} {\bibfnamefont {D.}~\bibnamefont {Pontoni}}, \ and\ \bibinfo
  {author} {\bibfnamefont {A.}~\bibnamefont {Stierle}},\ }\href@noop {}
  {\bibfield  {journal} {\bibinfo  {journal} {Nat. Comm.}\ }\textbf {\bibinfo
  {volume} {7}},\ \bibinfo {pages} {10964} (\bibinfo {year}
  {2016})}\BibitemShut {NoStop}%
\bibitem [{\citenamefont {Li}\ \emph {et~al.}(2012)\citenamefont {Li},
  \citenamefont {Nielsen}, \citenamefont {Lee}, \citenamefont {Frandsen},
  \citenamefont {Banfield},\ and\ \citenamefont {{De Yoreo}}}]{Li12}%
  \BibitemOpen
  \bibfield  {author} {\bibinfo {author} {\bibfnamefont {D.}~\bibnamefont
  {Li}}, \bibinfo {author} {\bibfnamefont {M.~H.}\ \bibnamefont {Nielsen}},
  \bibinfo {author} {\bibfnamefont {J.~R.~I.}\ \bibnamefont {Lee}}, \bibinfo
  {author} {\bibfnamefont {C.}~\bibnamefont {Frandsen}}, \bibinfo {author}
  {\bibfnamefont {J.~F.}\ \bibnamefont {Banfield}}, \ and\ \bibinfo {author}
  {\bibfnamefont {J.~J.}\ \bibnamefont {{De Yoreo}}},\ }\href@noop {}
  {\bibfield  {journal} {\bibinfo  {journal} {Science}\ }\textbf {\bibinfo
  {volume} {336}},\ \bibinfo {pages} {1014} (\bibinfo {year}
  {2012})}\BibitemShut {NoStop}%
\bibitem [{\citenamefont {Rauscher}\ \emph {et~al.}(1999)\citenamefont
  {Rauscher}, \citenamefont {Paniago}, \citenamefont {Metzger}, \citenamefont
  {Kovats}, \citenamefont {Domke}, \citenamefont {Peisl}, \citenamefont
  {Pfannes}, \citenamefont {Schulze},\ and\ \citenamefont
  {Eisele}}]{Rauscher99}%
  \BibitemOpen
  \bibfield  {author} {\bibinfo {author} {\bibfnamefont {M.}~\bibnamefont
  {Rauscher}}, \bibinfo {author} {\bibfnamefont {R.}~\bibnamefont {Paniago}},
  \bibinfo {author} {\bibfnamefont {H.}~\bibnamefont {Metzger}}, \bibinfo
  {author} {\bibfnamefont {Z.}~\bibnamefont {Kovats}}, \bibinfo {author}
  {\bibfnamefont {J.}~\bibnamefont {Domke}}, \bibinfo {author} {\bibfnamefont
  {J.}~\bibnamefont {Peisl}}, \bibinfo {author} {\bibfnamefont {H.-D.}\
  \bibnamefont {Pfannes}}, \bibinfo {author} {\bibfnamefont {J.}~\bibnamefont
  {Schulze}}, \ and\ \bibinfo {author} {\bibfnamefont {I.}~\bibnamefont
  {Eisele}},\ }\href@noop {} {\bibfield  {journal} {\bibinfo  {journal} {J.
  Appl. Phys.}\ }\textbf {\bibinfo {volume} {86}},\ \bibinfo {pages} {6763}
  (\bibinfo {year} {1999})}\BibitemShut {NoStop}%
\bibitem [{\citenamefont {Park}\ \emph {et~al.}(2005)\citenamefont {Park},
  \citenamefont {Lee}, \citenamefont {Rai}, \citenamefont {Mukherjee},\ and\
  \citenamefont {Zachariah}}]{Park05}%
  \BibitemOpen
  \bibfield  {author} {\bibinfo {author} {\bibfnamefont {K.}~\bibnamefont
  {Park}}, \bibinfo {author} {\bibfnamefont {D.}~\bibnamefont {Lee}}, \bibinfo
  {author} {\bibfnamefont {A.}~\bibnamefont {Rai}}, \bibinfo {author}
  {\bibfnamefont {D.}~\bibnamefont {Mukherjee}}, \ and\ \bibinfo {author}
  {\bibfnamefont {M.~R.}\ \bibnamefont {Zachariah}},\ }\href@noop {} {\bibfield
   {journal} {\bibinfo  {journal} {J. Phys. Chem. B}\ }\textbf {\bibinfo
  {volume} {109}},\ \bibinfo {pages} {7290} (\bibinfo {year}
  {2005})}\BibitemShut {NoStop}%
\bibitem [{\citenamefont {Pecora}(2000)}]{Pecora00}%
  \BibitemOpen
  \bibfield  {author} {\bibinfo {author} {\bibfnamefont {R.}~\bibnamefont
  {Pecora}},\ }\href@noop {} {\bibfield  {journal} {\bibinfo  {journal} {J.
  Nanopart. Res.}\ }\textbf {\bibinfo {volume} {2}},\ \bibinfo {pages} {123}
  (\bibinfo {year} {2000})}\BibitemShut {NoStop}%
\bibitem [{\citenamefont {Juv\'{e}}\ \emph {et~al.}(2013)\citenamefont
  {Juv\'{e}}, \citenamefont {Cardinal}, \citenamefont {Lombardi}, \citenamefont
  {Crut}, \citenamefont {Maioli}, \citenamefont {P\'{e}rez-Juste},
  \citenamefont {Liz-Marz\'{a}n}, \citenamefont {{Del Fatti}},\ and\
  \citenamefont {Vall\'{e}e}}]{Juve13}%
  \BibitemOpen
  \bibfield  {author} {\bibinfo {author} {\bibfnamefont {V.}~\bibnamefont
  {Juv\'{e}}}, \bibinfo {author} {\bibfnamefont {M.~F.}\ \bibnamefont
  {Cardinal}}, \bibinfo {author} {\bibfnamefont {A.}~\bibnamefont {Lombardi}},
  \bibinfo {author} {\bibfnamefont {A.}~\bibnamefont {Crut}}, \bibinfo {author}
  {\bibfnamefont {P.}~\bibnamefont {Maioli}}, \bibinfo {author} {\bibfnamefont
  {J.}~\bibnamefont {P\'{e}rez-Juste}}, \bibinfo {author} {\bibfnamefont
  {L.~M.}\ \bibnamefont {Liz-Marz\'{a}n}}, \bibinfo {author} {\bibfnamefont
  {N.}~\bibnamefont {{Del Fatti}}}, \ and\ \bibinfo {author} {\bibfnamefont
  {F.}~\bibnamefont {Vall\'{e}e}},\ }\href@noop {} {\bibfield  {journal}
  {\bibinfo  {journal} {Nano Lett.}\ }\textbf {\bibinfo {volume} {13}},\
  \bibinfo {pages} {2234} (\bibinfo {year} {2013})}\BibitemShut {NoStop}%
\bibitem [{\citenamefont {Shukla}\ \emph {et~al.}(2003)\citenamefont {Shukla},
  \citenamefont {Liu}, \citenamefont {Jones},\ and\ \citenamefont
  {Weller}}]{Shukla03}%
  \BibitemOpen
  \bibfield  {author} {\bibinfo {author} {\bibfnamefont {N.}~\bibnamefont
  {Shukla}}, \bibinfo {author} {\bibfnamefont {C.}~\bibnamefont {Liu}},
  \bibinfo {author} {\bibfnamefont {P.~M.}\ \bibnamefont {Jones}}, \ and\
  \bibinfo {author} {\bibfnamefont {D.}~\bibnamefont {Weller}},\ }\href@noop {}
  {\bibfield  {journal} {\bibinfo  {journal} {J. Magn. Magn. Mater.}\ }\textbf
  {\bibinfo {volume} {266}},\ \bibinfo {pages} {178} (\bibinfo {year}
  {2003})}\BibitemShut {NoStop}%
\bibitem [{\citenamefont {Amendola}\ and\ \citenamefont
  {Meneghetti}(2009)}]{Amendola09}%
  \BibitemOpen
  \bibfield  {author} {\bibinfo {author} {\bibfnamefont {V.}~\bibnamefont
  {Amendola}}\ and\ \bibinfo {author} {\bibfnamefont {M.}~\bibnamefont
  {Meneghetti}},\ }\href@noop {} {\bibfield  {journal} {\bibinfo  {journal} {J.
  Phys. Chem. C}\ }\textbf {\bibinfo {volume} {113}},\ \bibinfo {pages} {4277}
  (\bibinfo {year} {2009})}\BibitemShut {NoStop}%
\bibitem [{\citenamefont {Prieto}\ \emph {et~al.}(2012)\citenamefont {Prieto},
  \citenamefont {Nistor}, \citenamefont {Nouneh}, \citenamefont {Oyama},
  \citenamefont {Abd-Lefdil},\ and\ \citenamefont {D\'{i}az}}]{Prieto12}%
  \BibitemOpen
  \bibfield  {author} {\bibinfo {author} {\bibfnamefont {P.}~\bibnamefont
  {Prieto}}, \bibinfo {author} {\bibfnamefont {V.}~\bibnamefont {Nistor}},
  \bibinfo {author} {\bibfnamefont {K.}~\bibnamefont {Nouneh}}, \bibinfo
  {author} {\bibfnamefont {M.}~\bibnamefont {Oyama}}, \bibinfo {author}
  {\bibfnamefont {M.}~\bibnamefont {Abd-Lefdil}}, \ and\ \bibinfo {author}
  {\bibfnamefont {R.}~\bibnamefont {D\'{i}az}},\ }\href@noop {} {\bibfield
  {journal} {\bibinfo  {journal} {Appl. Surf. Sci.}\ }\textbf {\bibinfo
  {volume} {258}},\ \bibinfo {pages} {8807} (\bibinfo {year}
  {2012})}\BibitemShut {NoStop}%
\bibitem [{\citenamefont {Wiesendanger}(2009)}]{Wiesendanger09}%
  \BibitemOpen
  \bibfield  {author} {\bibinfo {author} {\bibfnamefont {R.}~\bibnamefont
  {Wiesendanger}},\ }\href@noop {} {\bibfield  {journal} {\bibinfo  {journal}
  {Rev. Mod. Phys.}\ }\textbf {\bibinfo {volume} {81}},\ \bibinfo {pages}
  {1495} (\bibinfo {year} {2009})}\BibitemShut {NoStop}%
\bibitem [{\citenamefont {McCord}(2015)}]{McCord15}%
  \BibitemOpen
  \bibfield  {author} {\bibinfo {author} {\bibfnamefont {J.}~\bibnamefont
  {McCord}},\ }\href@noop {} {\bibfield  {journal} {\bibinfo  {journal} {J.
  Phys. D: Appl. Phys.}\ }\textbf {\bibinfo {volume} {48}},\ \bibinfo {pages}
  {333001} (\bibinfo {year} {2015})}\BibitemShut {NoStop}%
\bibitem [{\citenamefont {Koike}(2013)}]{Koike13}%
  \BibitemOpen
  \bibfield  {author} {\bibinfo {author} {\bibfnamefont {K.}~\bibnamefont
  {Koike}},\ }\href@noop {} {\bibfield  {journal} {\bibinfo  {journal}
  {Microscopy}\ }\textbf {\bibinfo {volume} {62}},\ \bibinfo {pages} {177}
  (\bibinfo {year} {2013})}\BibitemShut {NoStop}%
\bibitem [{\citenamefont {Cheng}\ and\ \citenamefont
  {Keavney}(2012)}]{Cheng12}%
  \BibitemOpen
  \bibfield  {author} {\bibinfo {author} {\bibfnamefont {X.}~\bibnamefont
  {Cheng}}\ and\ \bibinfo {author} {\bibfnamefont {D.}~\bibnamefont
  {Keavney}},\ }\href@noop {} {\bibfield  {journal} {\bibinfo  {journal} {Rep.
  Prog. Phys.}\ }\textbf {\bibinfo {volume} {75}},\ \bibinfo {pages} {026501}
  (\bibinfo {year} {2012})}\BibitemShut {NoStop}%
\bibitem [{\citenamefont {Grimsditch}\ and\ \citenamefont
  {Vavassori}(2004)}]{Grimsditch04}%
  \BibitemOpen
  \bibfield  {author} {\bibinfo {author} {\bibfnamefont {M.}~\bibnamefont
  {Grimsditch}}\ and\ \bibinfo {author} {\bibfnamefont {P.}~\bibnamefont
  {Vavassori}},\ }\href@noop {} {\bibfield  {journal} {\bibinfo  {journal} {J.
  Phys.: Condens. Mat.}\ }\textbf {\bibinfo {volume} {16}},\ \bibinfo {pages}
  {R275} (\bibinfo {year} {2004})}\BibitemShut {NoStop}%
\bibitem [{\citenamefont {Hartmut}\ \emph {et~al.}(2007)\citenamefont
  {Hartmut}, \citenamefont {Theis-Br\"{o}hl},\ and\ \citenamefont
  {Toperverg}}]{Zabel07}%
  \BibitemOpen
  \bibfield  {author} {\bibinfo {author} {\bibfnamefont {H.}~\bibnamefont
  {Hartmut}}, \bibinfo {author} {\bibfnamefont {K.}~\bibnamefont
  {Theis-Br\"{o}hl}}, \ and\ \bibinfo {author} {\bibfnamefont {B.~P.}\
  \bibnamefont {Toperverg}},\ }\enquote {\bibinfo {title} {Polarized neutron
  reflectivity and scattering from magnetic nanostructures and spintronic
  materials},}\ in\ \href@noop {} {\emph {\bibinfo {booktitle} {Handbook of
  Magnetism and Advanced Magnetic Materials}}}\ (\bibinfo  {publisher} {John
  Wiley \& Sons, Ltd},\ \bibinfo {year} {2007})\BibitemShut {NoStop}%
\bibitem [{\citenamefont {St\"{o}hr}\ \emph {et~al.}(1993)\citenamefont
  {St\"{o}hr}, \citenamefont {Wu}, \citenamefont {Hermsmeier}, \citenamefont
  {Samant}, \citenamefont {Harp}, \citenamefont {Koranda}, \citenamefont
  {Dunham},\ and\ \citenamefont {Tonner}}]{Stoehr93}%
  \BibitemOpen
  \bibfield  {author} {\bibinfo {author} {\bibfnamefont {J.}~\bibnamefont
  {St\"{o}hr}}, \bibinfo {author} {\bibfnamefont {Y.}~\bibnamefont {Wu}},
  \bibinfo {author} {\bibfnamefont {B.}~\bibnamefont {Hermsmeier}}, \bibinfo
  {author} {\bibfnamefont {M.}~\bibnamefont {Samant}}, \bibinfo {author}
  {\bibfnamefont {G.}~\bibnamefont {Harp}}, \bibinfo {author} {\bibfnamefont
  {S.}~\bibnamefont {Koranda}}, \bibinfo {author} {\bibfnamefont
  {D.}~\bibnamefont {Dunham}}, \ and\ \bibinfo {author} {\bibfnamefont
  {B.}~\bibnamefont {Tonner}},\ }\href@noop {} {\bibfield  {journal} {\bibinfo
  {journal} {{Science}}\ }\textbf {\bibinfo {volume} {259}},\ \bibinfo {pages}
  {658} (\bibinfo {year} {1993})}\BibitemShut {NoStop}%
\bibitem [{\citenamefont {Turner}\ \emph {et~al.}(2011)\citenamefont {Turner},
  \citenamefont {Huang}, \citenamefont {Krupin}, \citenamefont {Seu},
  \citenamefont {Parks}, \citenamefont {Kevan}, \citenamefont {Lima},
  \citenamefont {Kisslinger}, \citenamefont {McNulty}, \citenamefont {Gambino},
  \citenamefont {Mangin}, \citenamefont {Roy},\ and\ \citenamefont
  {Fischer}}]{Turner11}%
  \BibitemOpen
  \bibfield  {author} {\bibinfo {author} {\bibfnamefont {J.}~\bibnamefont
  {Turner}}, \bibinfo {author} {\bibfnamefont {X.}~\bibnamefont {Huang}},
  \bibinfo {author} {\bibfnamefont {O.}~\bibnamefont {Krupin}}, \bibinfo
  {author} {\bibfnamefont {K.}~\bibnamefont {Seu}}, \bibinfo {author}
  {\bibfnamefont {D.}~\bibnamefont {Parks}}, \bibinfo {author} {\bibfnamefont
  {S.}~\bibnamefont {Kevan}}, \bibinfo {author} {\bibfnamefont
  {E.}~\bibnamefont {Lima}}, \bibinfo {author} {\bibfnamefont {K.}~\bibnamefont
  {Kisslinger}}, \bibinfo {author} {\bibfnamefont {I.}~\bibnamefont {McNulty}},
  \bibinfo {author} {\bibfnamefont {R.}~\bibnamefont {Gambino}}, \bibinfo
  {author} {\bibfnamefont {S.}~\bibnamefont {Mangin}}, \bibinfo {author}
  {\bibfnamefont {S.}~\bibnamefont {Roy}}, \ and\ \bibinfo {author}
  {\bibfnamefont {P.}~\bibnamefont {Fischer}},\ }\href@noop {} {\bibfield
  {journal} {\bibinfo  {journal} {Phys. Rev. Lett.}\ }\textbf {\bibinfo
  {volume} {107}},\ \bibinfo {pages} {033904} (\bibinfo {year}
  {2011})}\BibitemShut {NoStop}%
\bibitem [{\citenamefont {R\"{o}hlsberger}(1999)}]{Roehlsberger99}%
  \BibitemOpen
  \bibfield  {author} {\bibinfo {author} {\bibfnamefont {R.}~\bibnamefont
  {R\"{o}hlsberger}},\ }\href@noop {} {\bibfield  {journal} {\bibinfo
  {journal} {Hyperfine Interact.}\ }\textbf {\bibinfo {volume} {123/124}},\
  \bibinfo {pages} {455} (\bibinfo {year} {1999})}\BibitemShut {NoStop}%
\bibitem [{\citenamefont {R\"{o}hlsberger}\ \emph {et~al.}(2003)\citenamefont
  {R\"{o}hlsberger}, \citenamefont {Bansmann}, \citenamefont {Senz},
  \citenamefont {Jonas}, \citenamefont {Bettac}, \citenamefont {Meiwes-Broer},\
  and\ \citenamefont {Leupold}}]{Roehlsberger03}%
  \BibitemOpen
  \bibfield  {author} {\bibinfo {author} {\bibfnamefont {R.}~\bibnamefont
  {R\"{o}hlsberger}}, \bibinfo {author} {\bibfnamefont {J.}~\bibnamefont
  {Bansmann}}, \bibinfo {author} {\bibfnamefont {V.}~\bibnamefont {Senz}},
  \bibinfo {author} {\bibfnamefont {K.~L.}\ \bibnamefont {Jonas}}, \bibinfo
  {author} {\bibfnamefont {A.}~\bibnamefont {Bettac}}, \bibinfo {author}
  {\bibfnamefont {K.~H.}\ \bibnamefont {Meiwes-Broer}}, \ and\ \bibinfo
  {author} {\bibfnamefont {O.}~\bibnamefont {Leupold}},\ }\href@noop {}
  {\bibfield  {journal} {\bibinfo  {journal} {Phys. Rev. B}\ }\textbf {\bibinfo
  {volume} {67}},\ \bibinfo {pages} {245412} (\bibinfo {year}
  {2003})}\BibitemShut {NoStop}%
\bibitem [{\citenamefont {Schlage}\ and\ \citenamefont
  {R\"{o}hlsberger}(2013)}]{Schlage13}%
  \BibitemOpen
  \bibfield  {author} {\bibinfo {author} {\bibfnamefont {K.}~\bibnamefont
  {Schlage}}\ and\ \bibinfo {author} {\bibfnamefont {R.}~\bibnamefont
  {R\"{o}hlsberger}},\ }\href@noop {} {\bibfield  {journal} {\bibinfo
  {journal} {J. Elect. Spect. Rel. Phenom.}\ }\textbf {\bibinfo {volume}
  {189}},\ \bibinfo {pages} {187} (\bibinfo {year} {2013})}\BibitemShut
  {NoStop}%
\bibitem [{\citenamefont {Bocklage}\ \emph {et~al.}(2015)\citenamefont
  {Bocklage}, \citenamefont {Swoboda}, \citenamefont {Schlage}, \citenamefont
  {Wille}, \citenamefont {Dzemiantsova}, \citenamefont {Bajt}, \citenamefont
  {Meier},\ and\ \citenamefont {R\"{o}hlsberger}}]{Bocklage15}%
  \BibitemOpen
  \bibfield  {author} {\bibinfo {author} {\bibfnamefont {L.}~\bibnamefont
  {Bocklage}}, \bibinfo {author} {\bibfnamefont {C.}~\bibnamefont {Swoboda}},
  \bibinfo {author} {\bibfnamefont {K.}~\bibnamefont {Schlage}}, \bibinfo
  {author} {\bibfnamefont {H.-C.}\ \bibnamefont {Wille}}, \bibinfo {author}
  {\bibfnamefont {L.}~\bibnamefont {Dzemiantsova}}, \bibinfo {author}
  {\bibfnamefont {S.}~\bibnamefont {Bajt}}, \bibinfo {author} {\bibfnamefont
  {G.}~\bibnamefont {Meier}}, \ and\ \bibinfo {author} {\bibfnamefont
  {R.}~\bibnamefont {R\"{o}hlsberger}},\ }\href@noop {} {\bibfield  {journal}
  {\bibinfo  {journal} {Phys. Rev. Lett.}\ }\textbf {\bibinfo {volume} {114}},\
  \bibinfo {pages} {147601} (\bibinfo {year} {2015})}\BibitemShut {NoStop}%
\bibitem [{\citenamefont {R\"{o}hlsberger}(2004)}]{Roehlsberger04}%
  \BibitemOpen
  \bibfield  {author} {\bibinfo {author} {\bibfnamefont {R.}~\bibnamefont
  {R\"{o}hlsberger}},\ }\href@noop {} {\emph {\bibinfo {title} {{Springer
  Tracts in Modern Physics, Vol. 208: Nuclear Condensed Matter Physics with
  Synchrotron Radiation}}}}\ (\bibinfo  {publisher} {Springer, Berlin},\
  \bibinfo {year} {2004})\BibitemShut {NoStop}%
\bibitem [{\citenamefont {Schlage}\ \emph {et~al.}(2012)\citenamefont
  {Schlage}, \citenamefont {Couet}, \citenamefont {Roth}, \citenamefont
  {Vainio}, \citenamefont {R\"{u}ffer}, \citenamefont {Kashem}, \citenamefont
  {M\"{u}ller-Buschbaum},\ and\ \citenamefont {R\"{o}hlsberger}}]{Schlage12}%
  \BibitemOpen
  \bibfield  {author} {\bibinfo {author} {\bibfnamefont {K.}~\bibnamefont
  {Schlage}}, \bibinfo {author} {\bibfnamefont {S.}~\bibnamefont {Couet}},
  \bibinfo {author} {\bibfnamefont {S.~V.}\ \bibnamefont {Roth}}, \bibinfo
  {author} {\bibfnamefont {U.}~\bibnamefont {Vainio}}, \bibinfo {author}
  {\bibfnamefont {R.}~\bibnamefont {R\"{u}ffer}}, \bibinfo {author}
  {\bibfnamefont {M.~M.~A.}\ \bibnamefont {Kashem}}, \bibinfo {author}
  {\bibfnamefont {P.}~\bibnamefont {M\"{u}ller-Buschbaum}}, \ and\ \bibinfo
  {author} {\bibfnamefont {R.}~\bibnamefont {R\"{o}hlsberger}},\ }\href@noop {}
  {\bibfield  {journal} {\bibinfo  {journal} {New J. Phys.}\ }\textbf {\bibinfo
  {volume} {14}},\ \bibinfo {pages} {043007} (\bibinfo {year}
  {2012})}\BibitemShut {NoStop}%
\bibitem [{\citenamefont {Sharma}\ \emph {et~al.}(2015)\citenamefont {Sharma},
  \citenamefont {Gupta}, \citenamefont {Gupta}, \citenamefont {Schlage},\ and\
  \citenamefont {Wille}}]{Sharma15}%
  \BibitemOpen
  \bibfield  {author} {\bibinfo {author} {\bibfnamefont {G.}~\bibnamefont
  {Sharma}}, \bibinfo {author} {\bibfnamefont {A.}~\bibnamefont {Gupta}},
  \bibinfo {author} {\bibfnamefont {M.}~\bibnamefont {Gupta}}, \bibinfo
  {author} {\bibfnamefont {K.}~\bibnamefont {Schlage}}, \ and\ \bibinfo
  {author} {\bibfnamefont {H.-C.}\ \bibnamefont {Wille}},\ }\href@noop {}
  {\bibfield  {journal} {\bibinfo  {journal} {Phys. Rev. B}\ }\textbf {\bibinfo
  {volume} {92}},\ \bibinfo {pages} {224403} (\bibinfo {year}
  {2015})}\BibitemShut {NoStop}%
\bibitem [{\citenamefont {Vogl}\ \emph {et~al.}(2009)\citenamefont {Vogl},
  \citenamefont {Partyka-Jankowska}, \citenamefont {Zaj\k{a}c},\ and\
  \citenamefont {Chumakov}}]{Vogl09}%
  \BibitemOpen
  \bibfield  {author} {\bibinfo {author} {\bibfnamefont {G.}~\bibnamefont
  {Vogl}}, \bibinfo {author} {\bibfnamefont {E.}~\bibnamefont
  {Partyka-Jankowska}}, \bibinfo {author} {\bibfnamefont {M.}~\bibnamefont
  {Zaj\k{a}c}}, \ and\ \bibinfo {author} {\bibfnamefont {A.~I.}\ \bibnamefont
  {Chumakov}},\ }\href@noop {} {\bibfield  {journal} {\bibinfo  {journal}
  {Phys. Rev. B}\ }\textbf {\bibinfo {volume} {80}},\ \bibinfo {pages} {115406}
  (\bibinfo {year} {2009})}\BibitemShut {NoStop}%
\bibitem [{\citenamefont {Potapkin}\ \emph {et~al.}(2013)\citenamefont
  {Potapkin}, \citenamefont {McCammon}, \citenamefont {Glazyrin}, \citenamefont
  {Kantor}, \citenamefont {Kupenko}, \citenamefont {Prescher}, \citenamefont
  {Sinmyo}, \citenamefont {Smirnov}, \citenamefont {Chumakov}, \citenamefont
  {R\"{u}ffer},\ and\ \citenamefont {Dubrovinski}}]{Potapkin13}%
  \BibitemOpen
  \bibfield  {author} {\bibinfo {author} {\bibfnamefont {V.}~\bibnamefont
  {Potapkin}}, \bibinfo {author} {\bibfnamefont {C.}~\bibnamefont {McCammon}},
  \bibinfo {author} {\bibfnamefont {K.}~\bibnamefont {Glazyrin}}, \bibinfo
  {author} {\bibfnamefont {A.}~\bibnamefont {Kantor}}, \bibinfo {author}
  {\bibfnamefont {I.}~\bibnamefont {Kupenko}}, \bibinfo {author} {\bibfnamefont
  {C.}~\bibnamefont {Prescher}}, \bibinfo {author} {\bibfnamefont
  {R.}~\bibnamefont {Sinmyo}}, \bibinfo {author} {\bibfnamefont {G.~V.}\
  \bibnamefont {Smirnov}}, \bibinfo {author} {\bibfnamefont {A.~I.}\
  \bibnamefont {Chumakov}}, \bibinfo {author} {\bibfnamefont {R.}~\bibnamefont
  {R\"{u}ffer}}, \ and\ \bibinfo {author} {\bibfnamefont {L.}~\bibnamefont
  {Dubrovinski}},\ }\href@noop {} {\bibfield  {journal} {\bibinfo  {journal}
  {Nature Communications}\ }\textbf {\bibinfo {volume} {4}},\ \bibinfo {pages}
  {1427} (\bibinfo {year} {2013})}\BibitemShut {NoStop}%
\bibitem [{\citenamefont {Freindl}\ \emph {et~al.}(2013)\citenamefont
  {Freindl}, \citenamefont {Partyka-Jankowska}, \citenamefont {Kara\'s},
  \citenamefont {Zaj\k{a}c}, \citenamefont {Madej}, \citenamefont {Spiridis},
  \citenamefont {\'{S}l\k{e}zak}, \citenamefont {\'{S}l\k{e}zak}, \citenamefont
  {Wi\'{s}nios},\ and\ \citenamefont {Korecki}}]{Freindl13}%
  \BibitemOpen
  \bibfield  {author} {\bibinfo {author} {\bibfnamefont {K.}~\bibnamefont
  {Freindl}}, \bibinfo {author} {\bibfnamefont {E.}~\bibnamefont
  {Partyka-Jankowska}}, \bibinfo {author} {\bibfnamefont {W.}~\bibnamefont
  {Kara\'s}}, \bibinfo {author} {\bibfnamefont {M.}~\bibnamefont {Zaj\k{a}c}},
  \bibinfo {author} {\bibfnamefont {E.}~\bibnamefont {Madej}}, \bibinfo
  {author} {\bibfnamefont {N.}~\bibnamefont {Spiridis}}, \bibinfo {author}
  {\bibfnamefont {M.}~\bibnamefont {\'{S}l\k{e}zak}}, \bibinfo {author}
  {\bibfnamefont {T.}~\bibnamefont {\'{S}l\k{e}zak}}, \bibinfo {author}
  {\bibfnamefont {D.}~\bibnamefont {Wi\'{s}nios}}, \ and\ \bibinfo {author}
  {\bibfnamefont {J.}~\bibnamefont {Korecki}},\ }\href@noop {} {\bibfield
  {journal} {\bibinfo  {journal} {Surf. Sci.}\ }\textbf {\bibinfo {volume}
  {617}},\ \bibinfo {pages} {183} (\bibinfo {year} {2013})}\BibitemShut
  {NoStop}%
\bibitem [{\citenamefont {Chiu}\ \emph {et~al.}(2013)\citenamefont {Chiu},
  \citenamefont {Wu}, \citenamefont {Yao}, \citenamefont {Zhou}, \citenamefont
  {Zhang}, \citenamefont {Ozolins},\ and\ \citenamefont {Huang}}]{Chiu13}%
  \BibitemOpen
  \bibfield  {author} {\bibinfo {author} {\bibfnamefont {C.-Y.}\ \bibnamefont
  {Chiu}}, \bibinfo {author} {\bibfnamefont {H.}~\bibnamefont {Wu}}, \bibinfo
  {author} {\bibfnamefont {Z.}~\bibnamefont {Yao}}, \bibinfo {author}
  {\bibfnamefont {F.}~\bibnamefont {Zhou}}, \bibinfo {author} {\bibfnamefont
  {H.}~\bibnamefont {Zhang}}, \bibinfo {author} {\bibfnamefont
  {V.}~\bibnamefont {Ozolins}}, \ and\ \bibinfo {author} {\bibfnamefont
  {Y.}~\bibnamefont {Huang}},\ }\href@noop {} {\bibfield  {journal} {\bibinfo
  {journal} {JACS}\ }\textbf {\bibinfo {volume} {135}},\ \bibinfo {pages}
  {15489} (\bibinfo {year} {2013})}\BibitemShut {NoStop}%
\bibitem [{\citenamefont {Mdluli}\ \emph {et~al.}(2011)\citenamefont {Mdluli},
  \citenamefont {Sosibo}, \citenamefont {Mashazi}, \citenamefont {Nyokong},
  \citenamefont {Tshikhudo}, \citenamefont {Skepu},\ and\ \citenamefont {{van
  der Lingen}}}]{Mdluli11}%
  \BibitemOpen
  \bibfield  {author} {\bibinfo {author} {\bibfnamefont {P.~S.}\ \bibnamefont
  {Mdluli}}, \bibinfo {author} {\bibfnamefont {N.~M.}\ \bibnamefont {Sosibo}},
  \bibinfo {author} {\bibfnamefont {P.~N.}\ \bibnamefont {Mashazi}}, \bibinfo
  {author} {\bibfnamefont {T.}~\bibnamefont {Nyokong}}, \bibinfo {author}
  {\bibfnamefont {R.~T.}\ \bibnamefont {Tshikhudo}}, \bibinfo {author}
  {\bibfnamefont {A.}~\bibnamefont {Skepu}}, \ and\ \bibinfo {author}
  {\bibfnamefont {E.}~\bibnamefont {{van der Lingen}}},\ }\href@noop {}
  {\bibfield  {journal} {\bibinfo  {journal} {J. Mol. Struct.}\ }\textbf
  {\bibinfo {volume} {1004}},\ \bibinfo {pages} {131} (\bibinfo {year}
  {2011})}\BibitemShut {NoStop}%
\bibitem [{\citenamefont {Sun}\ \emph {et~al.}(2012)\citenamefont {Sun},
  \citenamefont {Kong}, \citenamefont {You}, \citenamefont {Song},
  \citenamefont {Ding},\ and\ \citenamefont {Yang}}]{Sun12}%
  \BibitemOpen
  \bibfield  {author} {\bibinfo {author} {\bibfnamefont {S.}~\bibnamefont
  {Sun}}, \bibinfo {author} {\bibfnamefont {C.}~\bibnamefont {Kong}}, \bibinfo
  {author} {\bibfnamefont {H.}~\bibnamefont {You}}, \bibinfo {author}
  {\bibfnamefont {X.}~\bibnamefont {Song}}, \bibinfo {author} {\bibfnamefont
  {B.}~\bibnamefont {Ding}}, \ and\ \bibinfo {author} {\bibfnamefont
  {Z.}~\bibnamefont {Yang}},\ }\href@noop {} {\bibfield  {journal} {\bibinfo
  {journal} {Cryst. Eng. Comm.}\ }\textbf {\bibinfo {volume} {14}},\ \bibinfo
  {pages} {40} (\bibinfo {year} {2012})}\BibitemShut {NoStop}%
\bibitem [{\citenamefont {Liu}(2011)}]{Liu11}%
  \BibitemOpen
  \bibfield  {author} {\bibinfo {author} {\bibfnamefont {X.-W.}\ \bibnamefont
  {Liu}},\ }\href@noop {} {\bibfield  {journal} {\bibinfo  {journal}
  {Langmuir}\ }\textbf {\bibinfo {volume} {27}},\ \bibinfo {pages} {9100}
  (\bibinfo {year} {2011})}\BibitemShut {NoStop}%
\bibitem [{\citenamefont {Mankin}\ \emph {et~al.}(2015)\citenamefont {Mankin},
  \citenamefont {Day}, \citenamefont {Gao}, \citenamefont {No}, \citenamefont
  {Kim}, \citenamefont {McClelland}, \citenamefont {Bell}, \citenamefont
  {Park},\ and\ \citenamefont {Lieber}}]{Mankin15}%
  \BibitemOpen
  \bibfield  {author} {\bibinfo {author} {\bibfnamefont {M.~N.}\ \bibnamefont
  {Mankin}}, \bibinfo {author} {\bibfnamefont {R.~W.}\ \bibnamefont {Day}},
  \bibinfo {author} {\bibfnamefont {R.}~\bibnamefont {Gao}}, \bibinfo {author}
  {\bibfnamefont {Y.-S.}\ \bibnamefont {No}}, \bibinfo {author} {\bibfnamefont
  {S.-K.}\ \bibnamefont {Kim}}, \bibinfo {author} {\bibfnamefont {A.~A.}\
  \bibnamefont {McClelland}}, \bibinfo {author} {\bibfnamefont {D.~C.}\
  \bibnamefont {Bell}}, \bibinfo {author} {\bibfnamefont {H.-G.}\ \bibnamefont
  {Park}}, \ and\ \bibinfo {author} {\bibfnamefont {C.~M.}\ \bibnamefont
  {Lieber}},\ }\href@noop {} {\bibfield  {journal} {\bibinfo  {journal} {Nano
  Lett.}\ }\textbf {\bibinfo {volume} {15}},\ \bibinfo {pages} {4776} (\bibinfo
  {year} {2015})}\BibitemShut {NoStop}%
\bibitem [{\citenamefont {Harn}\ \emph {et~al.}(2015)\citenamefont {Harn},
  \citenamefont {Yang}, \citenamefont {Tang}, \citenamefont {Chen},\ and\
  \citenamefont {Wu}}]{Harn15}%
  \BibitemOpen
  \bibfield  {author} {\bibinfo {author} {\bibfnamefont {Y.-W.}\ \bibnamefont
  {Harn}}, \bibinfo {author} {\bibfnamefont {T.-H.}\ \bibnamefont {Yang}},
  \bibinfo {author} {\bibfnamefont {T.-Y.}\ \bibnamefont {Tang}}, \bibinfo
  {author} {\bibfnamefont {M.-C.}\ \bibnamefont {Chen}}, \ and\ \bibinfo
  {author} {\bibfnamefont {J.-M.}\ \bibnamefont {Wu}},\ }\href@noop {}
  {\bibfield  {journal} {\bibinfo  {journal} {Chem. Cat. Chem.}\ }\textbf
  {\bibinfo {volume} {7}},\ \bibinfo {pages} {80} (\bibinfo {year}
  {2015})}\BibitemShut {NoStop}%
\bibitem [{\citenamefont {Li}\ \emph {et~al.}(2013)\citenamefont {Li},
  \citenamefont {Zhang}, \citenamefont {Wang}, \citenamefont {Yang},
  \citenamefont {Li}, \citenamefont {Zhu}, \citenamefont {Zhou}, \citenamefont
  {Han},\ and\ \citenamefont {Li}}]{Li13}%
  \BibitemOpen
  \bibfield  {author} {\bibinfo {author} {\bibfnamefont {R.}~\bibnamefont
  {Li}}, \bibinfo {author} {\bibfnamefont {F.}~\bibnamefont {Zhang}}, \bibinfo
  {author} {\bibfnamefont {D.}~\bibnamefont {Wang}}, \bibinfo {author}
  {\bibfnamefont {J.}~\bibnamefont {Yang}}, \bibinfo {author} {\bibfnamefont
  {M.}~\bibnamefont {Li}}, \bibinfo {author} {\bibfnamefont {J.}~\bibnamefont
  {Zhu}}, \bibinfo {author} {\bibfnamefont {X.}~\bibnamefont {Zhou}}, \bibinfo
  {author} {\bibfnamefont {H.}~\bibnamefont {Han}}, \ and\ \bibinfo {author}
  {\bibfnamefont {C.}~\bibnamefont {Li}},\ }\href@noop {} {\bibfield  {journal}
  {\bibinfo  {journal} {Nat. Comm.}\ }\textbf {\bibinfo {volume} {4}},\
  \bibinfo {pages} {1432} (\bibinfo {year} {2013})}\BibitemShut {NoStop}%
\bibitem [{\citenamefont {Nasirpouri}\ \emph {et~al.}(2011)\citenamefont
  {Nasirpouri}, \citenamefont {Bending}, \citenamefont {Peter},\ and\
  \citenamefont {Fang\-ohr}}]{Nasirpouri11}%
  \BibitemOpen
  \bibfield  {author} {\bibinfo {author} {\bibfnamefont {F.}~\bibnamefont
  {Nasirpouri}}, \bibinfo {author} {\bibfnamefont {S.~J.}\ \bibnamefont
  {Bending}}, \bibinfo {author} {\bibfnamefont {L.~M.}\ \bibnamefont {Peter}},
  \ and\ \bibinfo {author} {\bibfnamefont {H.}~\bibnamefont {Fang\-ohr}},\
  }\href@noop {} {\bibfield  {journal} {\bibinfo  {journal} {Thin Solid Films}\
  }\textbf {\bibinfo {volume} {519}},\ \bibinfo {pages} {8320} (\bibinfo {year}
  {2011})}\BibitemShut {NoStop}%
\bibitem [{\citenamefont {Disch}\ \emph {et~al.}(2011)\citenamefont {Disch},
  \citenamefont {Wetterskog}, \citenamefont {Hermann}, \citenamefont
  {Salazar-Alvarez}, \citenamefont {Busch}, \citenamefont {Br\"{u}ckel},
  \citenamefont {Bergstr\"{o}m},\ and\ \citenamefont {Kamali}}]{Disch11}%
  \BibitemOpen
  \bibfield  {author} {\bibinfo {author} {\bibfnamefont {S.}~\bibnamefont
  {Disch}}, \bibinfo {author} {\bibfnamefont {E.}~\bibnamefont {Wetterskog}},
  \bibinfo {author} {\bibfnamefont {R.~P.}\ \bibnamefont {Hermann}}, \bibinfo
  {author} {\bibfnamefont {G.}~\bibnamefont {Salazar-Alvarez}}, \bibinfo
  {author} {\bibfnamefont {P.}~\bibnamefont {Busch}}, \bibinfo {author}
  {\bibfnamefont {T.}~\bibnamefont {Br\"{u}ckel}}, \bibinfo {author}
  {\bibfnamefont {L.}~\bibnamefont {Bergstr\"{o}m}}, \ and\ \bibinfo {author}
  {\bibfnamefont {S.}~\bibnamefont {Kamali}},\ }\href@noop {} {\bibfield
  {journal} {\bibinfo  {journal} {Nano Lett.}\ }\textbf {\bibinfo {volume}
  {11}},\ \bibinfo {pages} {1651} (\bibinfo {year} {2011})}\BibitemShut
  {NoStop}%
\bibitem [{\citenamefont {Chen}\ \emph {et~al.}(2013)\citenamefont {Chen},
  \citenamefont {Chiang},\ and\ \citenamefont {Wang}}]{Chen13}%
  \BibitemOpen
  \bibfield  {author} {\bibinfo {author} {\bibfnamefont {C.-J.}\ \bibnamefont
  {Chen}}, \bibinfo {author} {\bibfnamefont {R.-K.}\ \bibnamefont {Chiang}}, \
  and\ \bibinfo {author} {\bibfnamefont {S.-L.}\ \bibnamefont {Wang}},\
  }\href@noop {} {\bibfield  {journal} {\bibinfo  {journal} {Cryst. Eng.
  Comm.}\ }\textbf {\bibinfo {volume} {15}},\ \bibinfo {pages} {9161} (\bibinfo
  {year} {2013})}\BibitemShut {NoStop}%
\bibitem [{\citenamefont {Josten}\ \emph {et~al.}(2017)\citenamefont {Josten},
  \citenamefont {Wetterskog}, \citenamefont {Glavic}, \citenamefont {Boesecke},
  \citenamefont {Feoktystov}, \citenamefont {Brauweiler-Reuters}, \citenamefont
  {R\"{u}cker}, \citenamefont {Salazar-Alvarez}, \citenamefont {Br\"{u}ckel},\
  and\ \citenamefont {Bergstr\"{o}m}}]{Josten17}%
  \BibitemOpen
  \bibfield  {author} {\bibinfo {author} {\bibfnamefont {E.}~\bibnamefont
  {Josten}}, \bibinfo {author} {\bibfnamefont {E.}~\bibnamefont {Wetterskog}},
  \bibinfo {author} {\bibfnamefont {A.}~\bibnamefont {Glavic}}, \bibinfo
  {author} {\bibfnamefont {P.}~\bibnamefont {Boesecke}}, \bibinfo {author}
  {\bibfnamefont {A.}~\bibnamefont {Feoktystov}}, \bibinfo {author}
  {\bibfnamefont {E.}~\bibnamefont {Brauweiler-Reuters}}, \bibinfo {author}
  {\bibfnamefont {U.}~\bibnamefont {R\"{u}cker}}, \bibinfo {author}
  {\bibfnamefont {G.}~\bibnamefont {Salazar-Alvarez}}, \bibinfo {author}
  {\bibfnamefont {T.}~\bibnamefont {Br\"{u}ckel}}, \ and\ \bibinfo {author}
  {\bibfnamefont {L.}~\bibnamefont {Bergstr\"{o}m}},\ }\href@noop {} {\bibfield
   {journal} {\bibinfo  {journal} {Sci. Rep.}\ }\textbf {\bibinfo {volume}
  {7}},\ \bibinfo {pages} {2808} (\bibinfo {year} {2017})}\BibitemShut
  {NoStop}%
\bibitem [{Sup()}]{Supplement}%
  \BibitemOpen
  \href@noop {} {}\bibinfo {note} {See Supplemental Material for details on:
  experimental realization, NRS and GISAXS analysis, magnetic hysteresis and
  depth dependence of magnetization orientation.}\BibitemShut {Stop}%
\bibitem [{\citenamefont {Heffelfinger}\ and\ \citenamefont
  {Carter}(1997)}]{Heffelfinger97}%
  \BibitemOpen
  \bibfield  {author} {\bibinfo {author} {\bibfnamefont {J.~R.}\ \bibnamefont
  {Heffelfinger}}\ and\ \bibinfo {author} {\bibfnamefont {C.~B.}\ \bibnamefont
  {Carter}},\ }\href@noop {} {\bibfield  {journal} {\bibinfo  {journal} {Surf.
  Sci.}\ }\textbf {\bibinfo {volume} {389}},\ \bibinfo {pages} {188} (\bibinfo
  {year} {1997})}\BibitemShut {NoStop}%
\bibitem [{\citenamefont {Liedke}\ \emph {et~al.}(2013)\citenamefont {Liedke},
  \citenamefont {K\"{o}rner}, \citenamefont {Lenz}, \citenamefont {Fritzsche},
  \citenamefont {Ranjan}, \citenamefont {Keller}, \citenamefont
  {\v{C}i\v{z}m\'{a}r}, \citenamefont {Zvyagin}, \citenamefont {Facsko},
  \citenamefont {Potzger}, \citenamefont {Lindner},\ and\ \citenamefont
  {Fassbender}}]{Liedke13}%
  \BibitemOpen
  \bibfield  {author} {\bibinfo {author} {\bibfnamefont {M.~O.}\ \bibnamefont
  {Liedke}}, \bibinfo {author} {\bibfnamefont {M.}~\bibnamefont {K\"{o}rner}},
  \bibinfo {author} {\bibfnamefont {K.}~\bibnamefont {Lenz}}, \bibinfo {author}
  {\bibfnamefont {M.}~\bibnamefont {Fritzsche}}, \bibinfo {author}
  {\bibfnamefont {M.}~\bibnamefont {Ranjan}}, \bibinfo {author} {\bibfnamefont
  {A.}~\bibnamefont {Keller}}, \bibinfo {author} {\bibfnamefont
  {E.}~\bibnamefont {\v{C}i\v{z}m\'{a}r}}, \bibinfo {author} {\bibfnamefont
  {S.~A.}\ \bibnamefont {Zvyagin}}, \bibinfo {author} {\bibfnamefont
  {S.}~\bibnamefont {Facsko}}, \bibinfo {author} {\bibfnamefont
  {K.}~\bibnamefont {Potzger}}, \bibinfo {author} {\bibfnamefont
  {J.}~\bibnamefont {Lindner}}, \ and\ \bibinfo {author} {\bibfnamefont
  {J.}~\bibnamefont {Fassbender}},\ }\href@noop {} {\bibfield  {journal}
  {\bibinfo  {journal} {Phys. Rev. B}\ }\textbf {\bibinfo {volume} {87}},\
  \bibinfo {pages} {024424} (\bibinfo {year} {2013})}\BibitemShut {NoStop}%
\bibitem [{\citenamefont {Babonneau}(2010)}]{Babonneau10}%
  \BibitemOpen
  \bibfield  {author} {\bibinfo {author} {\bibfnamefont {D.}~\bibnamefont
  {Babonneau}},\ }\href@noop {} {\bibfield  {journal} {\bibinfo  {journal} {J.
  Appl. Cryst.}\ }\textbf {\bibinfo {volume} {43}},\ \bibinfo {pages} {929}
  (\bibinfo {year} {2010})}\BibitemShut {NoStop}%
\bibitem [{\citenamefont {Sturhahn}(2000)}]{Sturhahn00}%
  \BibitemOpen
  \bibfield  {author} {\bibinfo {author} {\bibfnamefont {W.}~\bibnamefont
  {Sturhahn}},\ }\href@noop {} {\bibfield  {journal} {\bibinfo  {journal}
  {Hyperfine Interactions}\ }\textbf {\bibinfo {volume} {125}},\ \bibinfo
  {pages} {149} (\bibinfo {year} {2000})}\BibitemShut {NoStop}%
\bibitem [{\citenamefont {Shiratsuchi}\ \emph {et~al.}(2004)\citenamefont
  {Shiratsuchi}, \citenamefont {Endo}, \citenamefont {Yamamoto}, \citenamefont
  {Li},\ and\ \citenamefont {Bader}}]{Shiratsuchi04}%
  \BibitemOpen
  \bibfield  {author} {\bibinfo {author} {\bibfnamefont {Y.}~\bibnamefont
  {Shiratsuchi}}, \bibinfo {author} {\bibfnamefont {Y.}~\bibnamefont {Endo}},
  \bibinfo {author} {\bibfnamefont {M.}~\bibnamefont {Yamamoto}}, \bibinfo
  {author} {\bibfnamefont {D.}~\bibnamefont {Li}}, \ and\ \bibinfo {author}
  {\bibfnamefont {S.~D.}\ \bibnamefont {Bader}},\ }\href@noop {} {\bibfield
  {journal} {\bibinfo  {journal} {J. Appl. Phys.}\ }\textbf {\bibinfo {volume}
  {95}},\ \bibinfo {pages} {6897} (\bibinfo {year} {2004})}\BibitemShut
  {NoStop}%
\end{thebibliography}
\end{document}


\newcommand{\AlOx}{$\alpha$-Al$_{2}$O$_{3}$}
\newcommand{\MFe}{$^{57}$Fe}
\newcommand{\dgr}{$^{\circ}$}

\renewcommand*{\citenumfont}[1]{S#1}
\renewcommand*{\bibnumfmt}[1]{[S#1]}
\renewcommand{\theequation}{S\arabic{equation}}
\renewcommand{\thefigure}{S\arabic{figure}}

\begin{center}
\LARGE{Supplemental material}
\end{center}

\title{Disentangling magnetic order on nanostructured surfaces}

\author{D. Erb}
\email[]{d.erb@hzdr.de}
\altaffiliation{present address: Institute of Ion Beam Physics and Materials Research, Helmholtz-Zentrum Dresden-Rossendorf, Bautzner Landstra{\ss}e 400, 01328 Dresden, Germany}
\affiliation{Deutsches Elektronen-Synchrotron DESY, Notkestra{\ss}e 85, 22607 Hamburg, Germany}

\author{K. Schlage}
\affiliation{Deutsches Elektronen-Synchrotron DESY, Notkestra{\ss}e 85, 22607 Hamburg, Germany}

\author{L. Bocklage}
\affiliation{Deutsches Elektronen-Synchrotron DESY, Notkestra{\ss}e 85, 22607 Hamburg, Germany}
\affiliation{The Hamburg Centre for Ultrafast Imaging, Luruper Chaussee 149, 22761 Hamburg, Germany}

\author{R. H\"{u}bner}
\affiliation{Institute of Ion Beam Physics and Materials Research, Helmholtz-Zentrum Dresden-Rossendorf, Bautzner Landstra{\ss}e 400, 01328 Dresden, Germany}

\author{D. G. Merkel}
\altaffiliation{on leave from Wigner Research Centre for Physics, Hungarian Academy of Sciences, H-1525 Budapest, Hungary}
\affiliation{European Synchrotron Radiation Facility, 71 avenue des Martyrs, 38000 Grenoble, France}

\author{R. R\"{u}ffer}
\affiliation{European Synchrotron Radiation Facility, 71 avenue des Martyrs, 38000 Grenoble, France}

\author{H.-C. Wille}
\affiliation{Deutsches Elektronen-Synchrotron DESY, Notkestra{\ss}e 85, 22607 Hamburg, Germany}

\author{R. R\"{o}hlsberger}
\affiliation{Deutsches Elektronen-Synchrotron DESY, Notkestra{\ss}e 85, 22607 Hamburg, Germany}
\affiliation{The Hamburg Centre for Ultrafast Imaging, Luruper Chaussee 149, 22761 Hamburg, Germany}

\date{\today}
\maketitle

\section{1. Author contributions}

K.S. conceived the experiments and developed the setup. K.S., D.E., and L.B. conducted the experiments. R.R\"{u}., D.G.M., and H.C.W. operated the beamlines and provided experimental support. D.E. and K.S. fitted the NRS data. D.E. simulated the GISAXS data. D.E. prepared and characterized the nanofaceted substrate. R.H. performed the TEM measurement. D.E. wrote the manuscript with K.S., L.B., and R.R\"{o}. All co-authors commented on the manuscript.

\section{2. Experimental setup}

\begin{figure}[h]
\begin{minipage}{0.60\textwidth}
\includegraphics[width=0.9\textwidth, trim={0cm, 0cm, 1.5cm, 0cm}, clip]{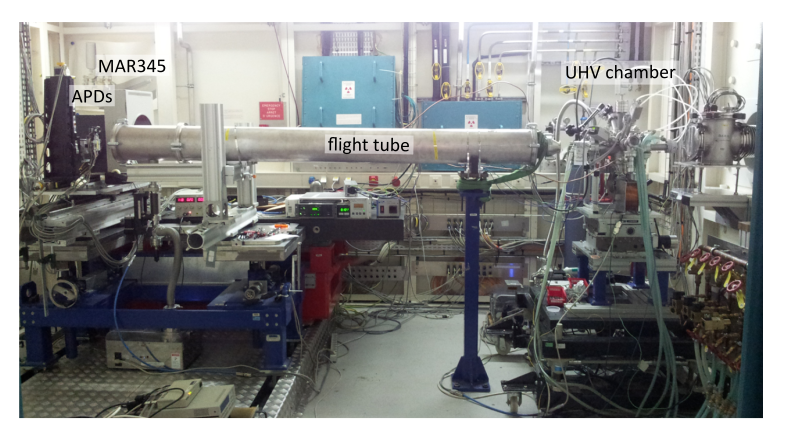}%
\end{minipage}
\hfill
\begin{minipage}{0.35\textwidth}
\caption{The GINSAXS experimental setup at the nuclear resonance beamline ID18 at ESRF. The UHV sputter deposition device is seen on the right, the detector stage on the left. An evacuated tube bridges the distance between sample and detectors to reduce scattering in air.}
\label{fig:FigureSM01}
\end{minipage}
\end{figure}

\begin{figure}[ht]
\begin{minipage}{0.60\textwidth}
\includegraphics[width=0.9\textwidth, trim={1.0cm, 1.75cm, 1.5cm, 0.75cm}, clip]{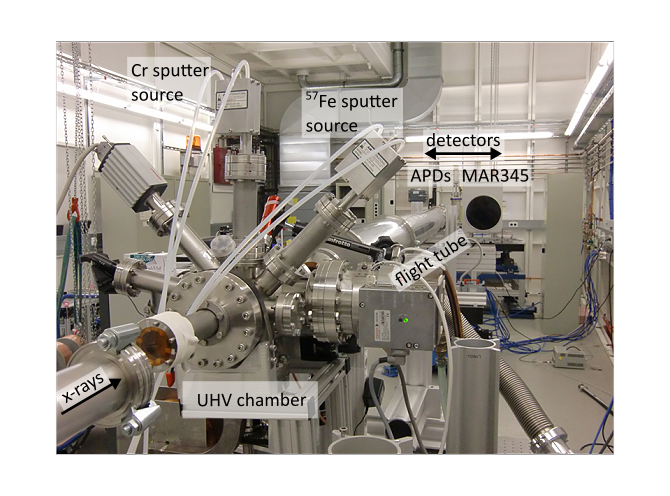}%
\end{minipage}
\hfill
\begin{minipage}{0.35\textwidth}
\caption{The GINSAXS experimental setup at the high resolution dynamics beamline P01 at PETRA III. The UHV sputter deposition device is seen in the foreground, the detector stage in the background. An evacuated tube bridges the distance between sample and detectors to reduce scattering in air.}
\label{fig:FigureSM02}
\end{minipage}
\end{figure}

The described experiments were carried out using a specialized setup, consisting mainly of a custom-made UHV sputter deposition device and a stage carrying two detectors (see Figs.~\ref{fig:FigureSM01} and \ref{fig:FigureSM02}).
The sputter deposition chamber is designed for in-situ GISAXS experiments. It is mobile and can be set up at any suitable x-ray beamline. The incident and scattered x-rays enter and exit the chamber via beryllium windows. The device features sample rotation around the surface normal (vertical; incidence angle and sample tilt are adjusted by means of an external double-tilt stage), sample heating, and application of an external magnetic field of up to 75~mT. Sputter sources can be installed at angles of 90\dgr, 45\dgr, 20\dgr, and 10\dgr~with respect to the sample surface. The base pressure of the UHV chamber was $3 \times 10^{-7}$~mbar; the pressure of the working gas Ar was $7.5 \times 10^{-3}$~mbar for \MFe~deposition. \MFe~was deposited stepwise at room temperature from the 45\dgr~source position, from a target of 1~inch in diameter at a power of 6~Watts. To reduce the angular divergence of the \MFe~atoms reaching the \AlOx~substrate, a horizontal slit mask was placed in front of the sputtering source.

For the ex-situ GINSAXS experiment, an external magnetic field higher than that achievable in the UHV chamber was required. Therefore, the UHV chamber was omitted in this experiment (the sample had been capped to prevent oxidation after the in-situ experiment) and the sample was placed between the pole shoes of an electromagnet mounted on a double-tilt stage. The field was applied in the sample plane, perpendicular to the orientation of the easy axis of magnetization and to the in-plane direction of the incoming x-ray beam, with a maximum strength of $\pm 300$~mT.

Both the area detector and the time-resolving point detector are mounted on a horizontal linear stage. Thus, area detector and point detector can be exchanged quickly after a deposition step or magnetic field step to record both the GISAXS pattern and the NRS time spectra for the respective state of the sample. A MAR345 image plate detector was employed to take the GISAXS patterns; the NRS time spectra were recorded by a stack of avalanche photo diodes (APDs).

The distance of 3010~mm between the sample position and the detector stage was bridged by an evacuated tube with Kapton windows to reduce scattering in air.

The NRS time spectra were recorded at an incidence angle of the incoming x-rays of $\alpha_{i} = 0.16$\dgr, i.e. close to the critical angle of the thin Fe film on \AlOx, where the nuclear resonant intensity is highest. The GISAXS patterns, however, had to be taken at an incidence angle of $\alpha_{i} = 0.6$\dgr, where the total scattered intensity is low enough to avoid overexposure and potential damage to the image plate.

\section{3. Substrate preparation}

\begin{wrapfigure}[22]{i}{10cm}
\vspace{-0.5cm}
\includegraphics[width=0.5\textwidth]{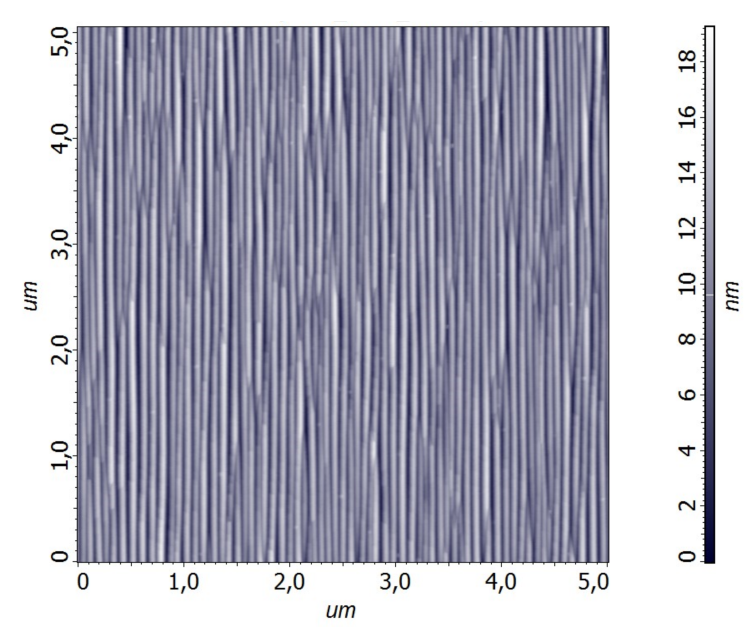}
\caption{Atomic force microscopy topography image of the nanofaceted \AlOx~ substrate after annealing in air at 1400~\dgr C for 24~h.}
\label{fig:FigureSM03}
\end{wrapfigure}

The nanofaceted substrate was prepared from a polished \AlOx wafer of 15~mm~$\times$~15~mm with M-plane $\left(10\bar{1}0\right)$ surface orientation. To induce the formation of surface facets \cite{Heffelfinger97}, the wafer was annealed at 1400~\dgr C in a high-temperature tube furnace in air for 24 hours. During annealing, the initial M-plane surface is reconstructed into nanoscale facets with R-plane $\left(1\bar{1}02\right)$ and S-plane $\left(10\bar{1}1\right)$ surface orientation. The facet edges are parallel to the $\left[11\bar{2}0\right]$ direction. The macroscopic surface orientation remains unchanged. Fig.~\ref{fig:FigureSM03} shows an atomic force microscopy topography image of the nanofaceted substrate: The average values for the geometric dimensions of the facets are obtained from atomic force microscopy (AFM) and grazing incidence small angle x-ray scattering (GISAXS): period $L = 80$~nm, width of the R-plane and S-plane facet surfaces $w_{R} = 32$~nm and $w_{S} = 55$~nm, height $h = 15$~nm, tilt angles $\beta_{R} = 30$\dgr~and $\beta_{R} = 17$\dgr.

\section{4. Details of NRS analysis}

The simulations of NRS time spectra were performed with the program CONUSS \cite{Sturhahn00}. In CONUSS, the orientation of the scattering plane is fixed with respect to an external frame of reference and serves as a reference plane for the orientations of the electric and magnetic field vectors and the magnetic hyperfine field. As this program was designed for analyzing time spectra of planar samples, the nanofaceted morphology of the \MFe~film on the \AlOx~substrate has to be accounted for. 

X-rays produced by a synchrotron are fully polarized with the electric (magnetic) field vector being parallel (perpendicular) to the plane of the storage ring, i.e. the horizontal plane. In the presented GINSAXS experiment, the macroscopic sample surface lies in the horizontal plane, but the tilts of the nanofacet surfaces with respect to the macroscopic sample surface result in two scattering planes which are tilted from the vertical plane and thus from the planes of the electric and the magnetic field vector. Having adjusted the input for the orientation of the electric field vector according to the different facet tilt angles of $\theta_{R} = 30$\dgr~and $\theta_{R} = 17$\dgr, the film regions on the R-plane and the S-plane facets can be treated as individual extended films in reflection geometry for fitting the respective time spectra.

\begin{figure}[hb]
\begin{minipage}{0.60\textwidth}
\includegraphics[width=\textwidth]{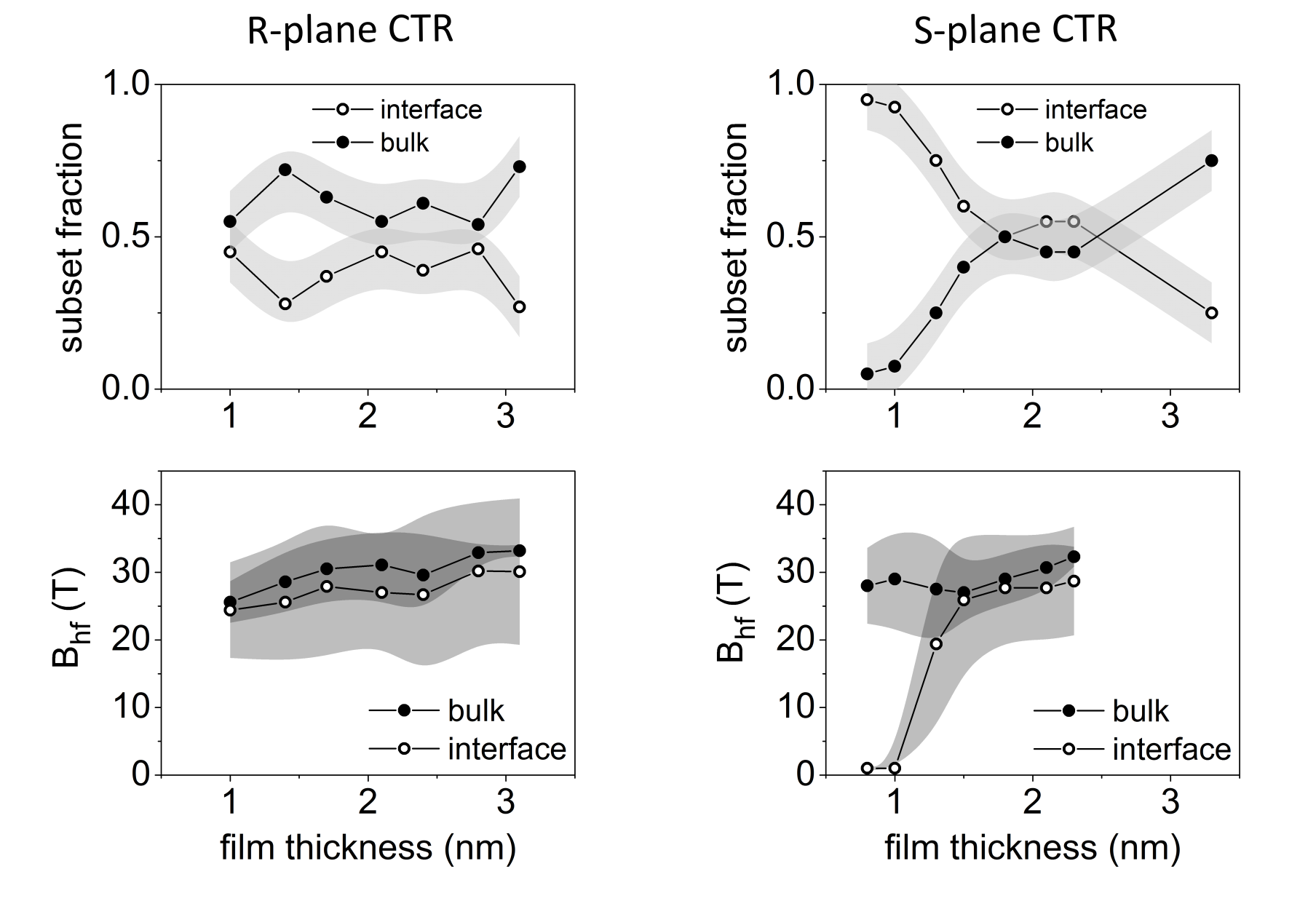}%
\end{minipage}
\hfill
\begin{minipage}{0.35\textwidth}
\caption{The evolution of the fractions of the bulk and interface components of \MFe~atoms in the sample (top) and of the magnetic hyperfine field strength and its distribution (bottom) for both thicker \MFe~film regions on the R-plane facets and thinner film regions on the S-plane facets during \MFe~deposition as extracted from fits of NRS time spectra recorded in situ.}
\label{fig:FigureSM04}
\end{minipage}
\end{figure}

CONUSS allows for defining subsets of resonant atoms with different properties. Here, two subsets Set1 and Set2 of \MFe~atoms were assumed for the simulations: Set1 is identified with \MFe~atoms at film and grain interfaces, while Set2 is associated with \MFe~atoms in a bulk-like atomic configuration. Time spectra recorded at the R-plane and S-plane CTR, i.e. for the thicker and the thinner \MFe~film regions, respectively, were fitted for all stages of the stepwise \MFe~deposition. From one deposition stage to the next, the magnetic hyperfine field strengths $B_{hf}$ and their respective distributions, as well as the weights of the two subsets were varied. The relative weights of subsets Set1 and Set2 in dependence of the respective film thickness are plotted in fig.~\ref{fig:FigureSM04}~(top). While no clear trend can be found for the subset weights in the thicker film regions, in the thinner film regions the fraction of Set1 atoms decreases as the volume fraction of interface atoms decreases with proceeding deposition. Fig.~\ref{fig:FigureSM04}~(bottom)shows the evolution of the magnetic hyperfine field strengths $B_{hf}$ of the different regions of the \MFe~film. The expected values of the Gaussian distribution of $B_{hf}$ are plotted against the increasing film thickness during deposition, with the shaded areas representing the full width at half maximum (FWHM) of the distribution.

For both the thicker and the thinner \MFe~film regions the fits show that the atoms of Set1 (interface component) are subject to a magnetic hyperfine field of strength $B_{hf}$, which is by approximately 10\% lower than that for by the Set2 (bulk component) atoms, and has a FWHM which is by a factor of about 2 larger than that for the Set2 atoms. The establishment of ferromagnetic behavior is indicated by a rapid decrease by a factor of 5 of the FWHM of the magnetic hyperfine field strength distribution seen by the bulk atoms (Set2) at film thicknesses of 2.8~nm for the thicker and at 2.3~nm for the thinner \MFe~film regions. It is peculiar, that the strength of the magnetic hyperfine fields does not increase monotonously with increasing film thickness. Instead, the strength of the magnetic hyperfine field decreases briefly, then increases again with reduced FWHM. Simulations assuming values for the magnetic hyperfine fields which conform to a monotonous increase do not yield adequate fits to the experimental data. This non-monotonous evolution of the magnetic hyperfine fields may be due to dependencies of the magnetic properties on several factors such as the growth mechanism of \MFe~on nanofaceted \AlOx~for the given incidence angles of sputtered atoms on the R-plane and S-plane facet surfaces, the resulting microstructure of the film, or the corrugated film shape with alternating stripe-like regions of different thickness. Due to the long duration of the experiment (24 hours), an influence of oxidation effects can be considered, too. Further investigations of the structure of the \MFe~film and its chemical composition would be required to gain more insight into the reasons why the magnetic hyperfine fields evolve in this way.

At 3.1~nm thickness, the film regions on the R-plane facet surfaces have a magnetic hyperfine field strength of $B_{hf} = 33.2$~T (in Set2), very close to the $\alpha$-\MFe~bulk value of $B_{hf} = 33.3$~T. At this deposition stage, the film regions on the S-plane facet surfaces have reached a thickness of 2.3~nm and a magnetic hyperfine field strength of $B_{hf} = 32.4$~T (in Set2). Both the thin and thick film regions now show time spectra shapes which are characteristic of ferromagnetically ordered \MFe~with the magnetization oriented parallel to the direction of the incoming beam: Due to the magnetic anisotropy induced by the uniaxially corrugated shape of the \MFe~film on the faceted substrate, the magnetization is oriented along the facet edges.

\section{5. Details of GISAXS simulations}

\begin{figure}[hb]%
\includegraphics[width=\linewidth, trim={0cm, 0.25cm, 0.75cm, 0cm}, clip]{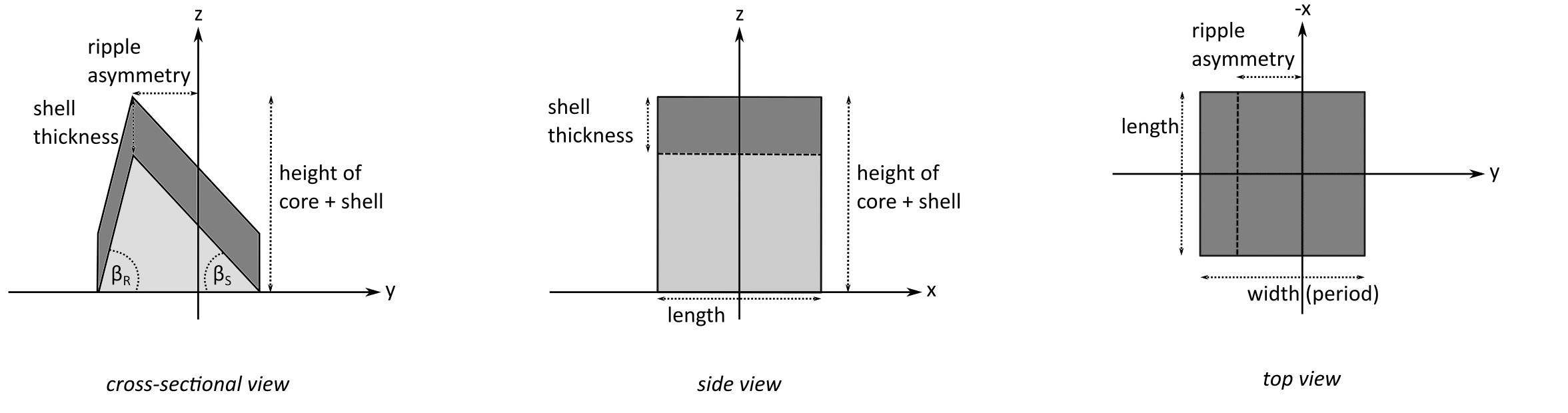}%
\caption{Illustration of the definition of the form factor ``core shell ripple'' in the program FitGISAXS \cite{Babonneau10} used to simulate the GISAXS patterns of the \MFe~film on faceted \AlOx.}%
\label{fig:FigureSM05}%
\end{figure}

GISAXS patterns were simulated using the software package FitGISAXS \cite{Babonneau10}. The form factor called ``core shell ripple'' (see Fig.~\ref{fig:FigureSM05}) was newly implemented by D. Babonneau to enable these simulations. Since the definition of the form factor did not allow choosing the thicknesses of the shell on the two surfaces of the facet independently, the left and right halves of the scattering patterns had to be simulated separately. Values for the optical constants of the sample constituents were taken from Ref.~\cite{CXRO} for an x-ray energy of 14.4~keV. Once the parameters describing the geometry of the substrate facets were set, only the \MFe~film thickness was varied to match the simulations with the experimental data. Due to the definition of the form factor, the thicknesses resulting from the simulations had to be multiplied by a factor of $\cos{(\beta_{R,S})}$, to obtain the film thicknesses as measured perpendicular to the respective facet surface.

\begin{figure}[h]
\includegraphics[width=\textwidth, trim={0cm, 0cm, 0cm, -0.75cm}]{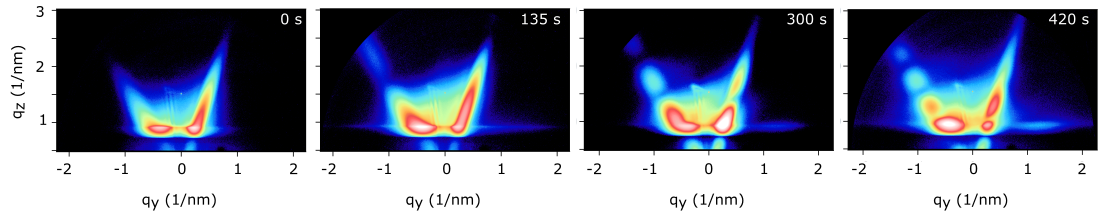}%
\caption{Sequence of selected GISAXS patterns of the \MFe~film on the nanofacetted \AlOx substrate. Labels state the respective duration of \MFe~deposition. The frequencies of the intensity modulations along the R-plane and S-plane crystal truncation rods correlates with the increasing film thicknesses on the respective facet faces.}
\label{fig:FigureSM06}
\end{figure}

\begin{figure}[h]
\begin{minipage}{0.60\textwidth}
\includegraphics[width=0.6\textwidth, trim={2cm, 0cm, 4cm, 0cm}, clip]{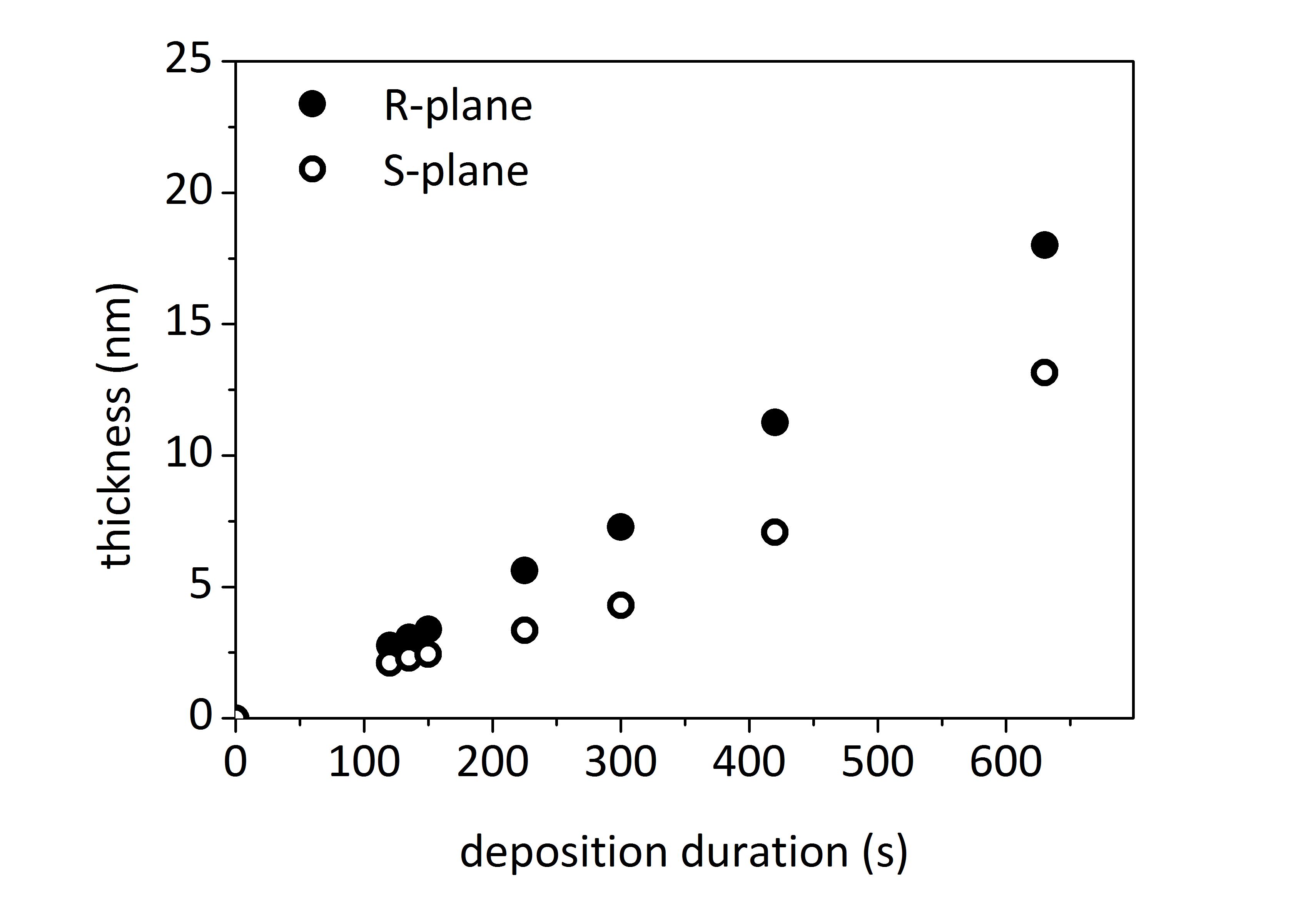}%
\end{minipage}
\hfill
\begin{minipage}{0.35\textwidth}
\caption{Evolution of the thicknesses of the \MFe~film on the R-plane and S-plane facets, respectively, of the \AlOx~substrate. The deposition rates are not constant due to the experimental conditions (see main text).}
\label{fig:FigureSM07}
\end{minipage}
\end{figure}

From simulations of the sequence of scattering patterns recorded during Fe deposition (see Fig.~\ref{fig:FigureSM06}), the evolution of the \MFe~film thickness on R-plane and S-plane facets was obtained, as plotted in Fig.~\ref{fig:FigureSM07}. Evidently, the depositions rates are not constant, but increase for later deposition stages. This is explained by the circumstances of the experiment: No external magnetic field was applied to the sample before 120~seconds of \MFe~deposition. After that, an external magnetic field was applied to the sample in several deposition stages to test the response of the magnetization to the external field. While the applied external magnetic field was too weak to cause any changes in magnetization, a residual magnetization of the pole shoes in the deposition chamber drastically influenced the plasma during sputter deposition, causing the \MFe~deposition rates to vary. The accessible angular range limited the number of detectable intensity oscillations along the CTRs, so that thicknesses below about 2.5~nm had to be extrapolated.

\begin{figure}[h]
\begin{minipage}{0.60\textwidth}
\includegraphics[width=0.75\textwidth, trim={0.5cm, 0cm, 1.5cm, -0.5cm}, clip]{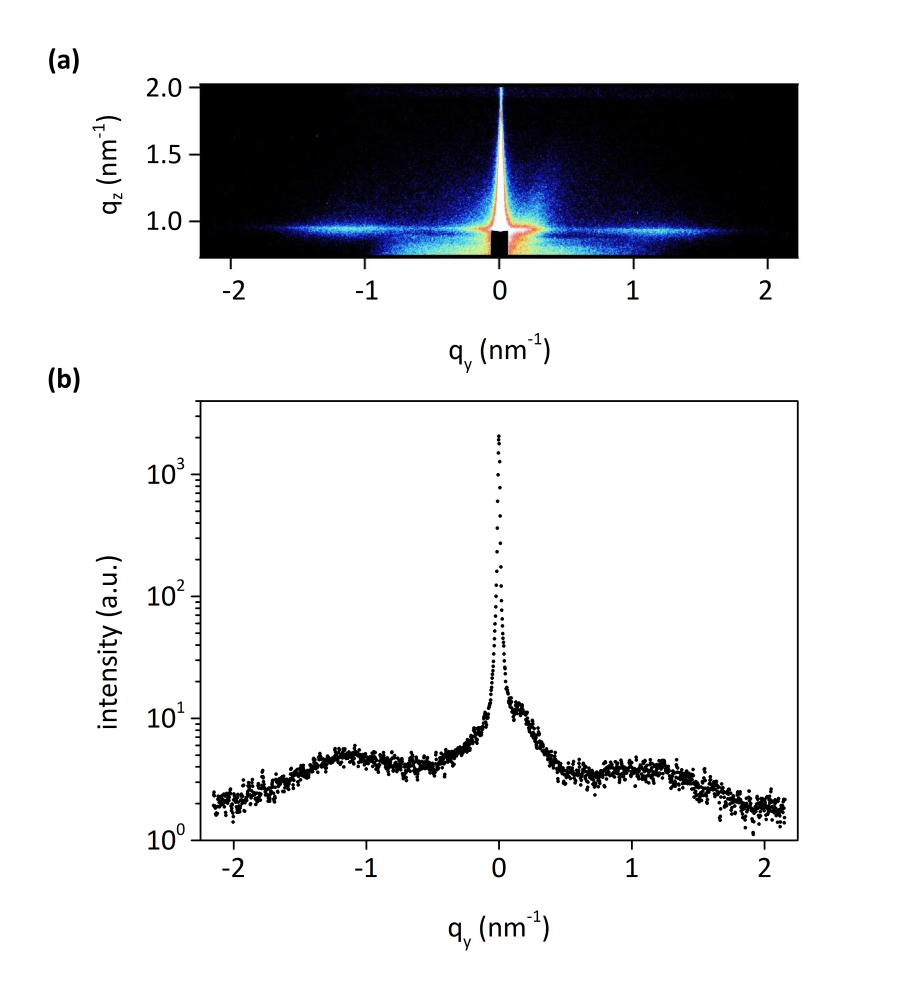}%
\end{minipage}
\hfill
\begin{minipage}{0.35\textwidth}
\caption{GISAXS data of the sample used for determining the crystallite size of the polycrystalline \MFe~film: (a) 2D scattering pattern and (b) horizontal section through (a).}
\label{fig:FigureSM08}
\end{minipage}
\end{figure}

Depositing \MFe~onto the nanofaceted \AlOx~substrate at room temperature resulted in a polycrystalline film. Fig.~\ref{fig:FigureSM08} shows GISAXS data of the sample oriented with the facet edges perpendicular to the direction of the incident X-ray beam. The positions of the broad off-specular intensity maxima at $q_{y} \approx \pm 1.2$~nm$^{-1}$ correspond to a lateral correlation length of approx. 5 nm, which is identified with the crystallite size of the film and agrees well with cross-sectional TEM data as shown in Fig.2(c) of the article.

\clearpage

\section{6. Magnetic hysteresis of the \MFe~film and depth dependence of the magnetization orientation}

\begin{figure}[h]
\begin{minipage}{0.60\textwidth}
\includegraphics[width=\textwidth, trim={0cm, 0cm, 0cm, 0cm}, clip]{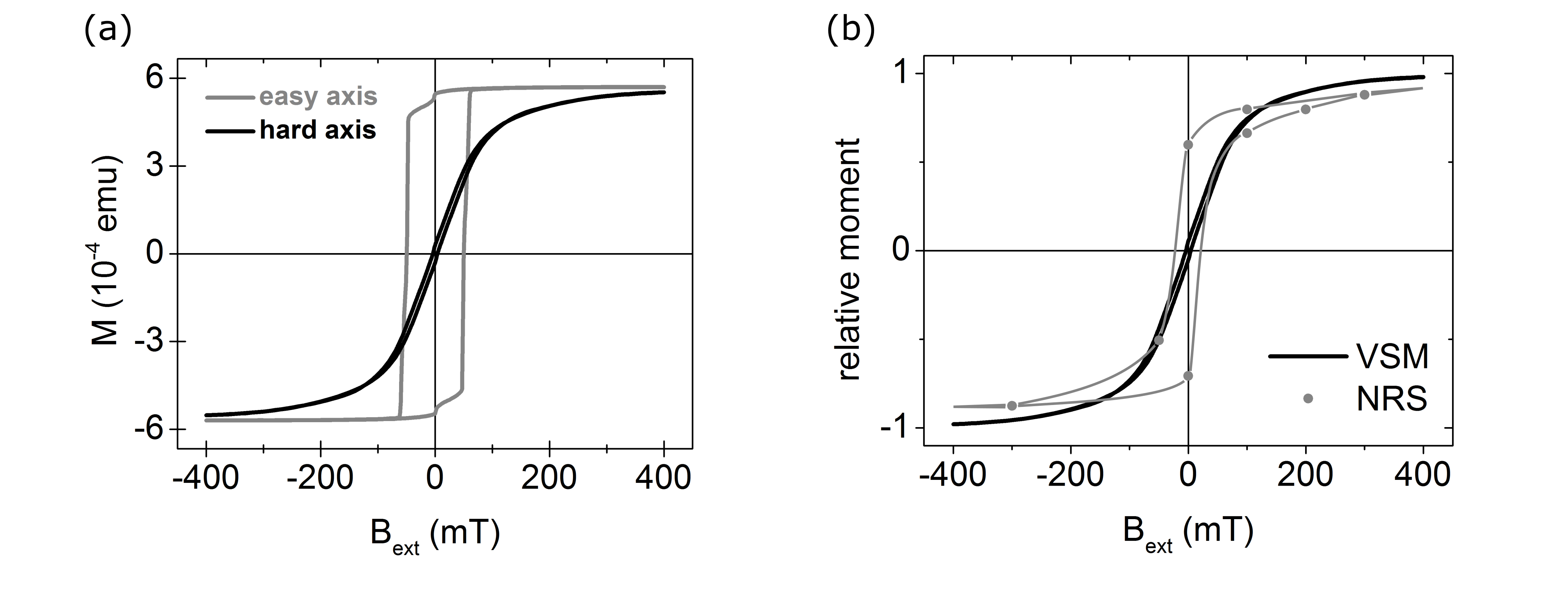}%
\end{minipage}
\hfill
\begin{minipage}{0.35\textwidth}
\caption{(a) Hysteresis loops of the Fe film on faceted \AlOx~with the external magnetic field parallel (gray) and perpendicular (black) to the facet edges, i.e. along the easy and the hard axis of magnetization. (b) Comparison of the magnetic hystereses as calculated from the magnetization orientations obtained from NRS at $\alpha_{i} = 0.16^{\circ}$ (gray symbols; lines are a guide to the eye) and as measured by VSM (black curve; hard axis).}
\label{fig:FigureSM09}
\end{minipage}
\end{figure}

Fig.~\ref{fig:FigureSM09}(a) shows hysteresis loops of the sample recorded via vibrating sample magnetometry (VSM) at 300~K with the external magnetic field applied parallel and perpendicular to the facet edges, respectively, evidencing the pronounced uniaxial magnetic anisotropy of the \MFe~film. 

While VSM measures the magnetic moment integrated over the entire sample volume, information obtained from NRS is depth-dependent: The incidence angle of the x-rays determines their penetration depth into the sample and thus the probed sample volume: with increasing incidence angle, the penetration depth increases. At $\alpha_{i} = 0.16^{\circ}$, as given in the presented NRS experiments, the x-rays penetrate only the few topmost monolayers of the Fe layer - the resonantly scattered intensity, however, is highest at this angle \cite{Roehlsberger04}. Fig.~\ref{fig:FigureSM09}(b) compares the hysteresis calculated from the magnetization orientations obtained from NRS with the hard axis hysteresis measured by VSM. Since NRS at $\alpha_{i} = 0.16^{\circ}$ probes only the topmost Fe monolayers, the discrepancies between the two hysteresis curves, i.e. the higher coercive field and lesser saturation moment, can be attributed to interface coupling between this part of the Fe film to the Cr capping layer.

\begin{figure}[h]
\begin{minipage}{0.60\textwidth}
\includegraphics[width=\textwidth, trim={0cm, 0cm, 0cm, 0cm}, clip]{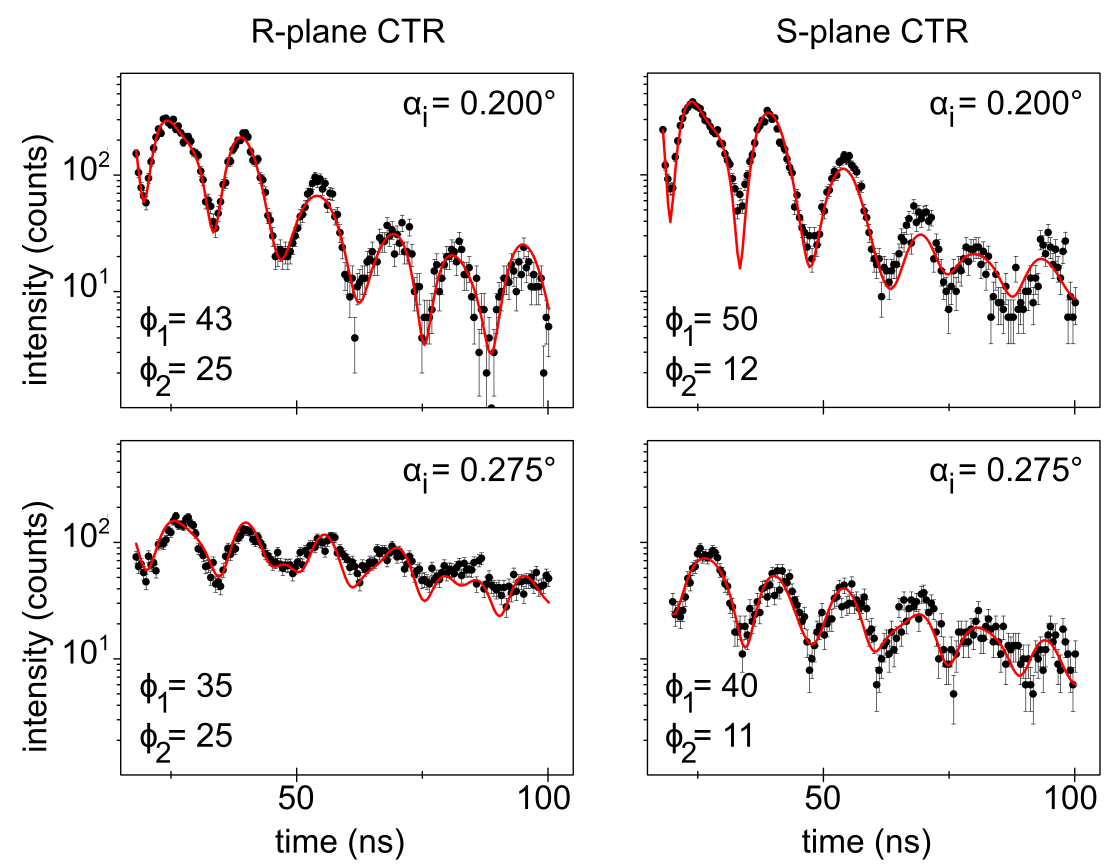}%
\end{minipage}
\hfill
\begin{minipage}{0.35\textwidth}
\caption{NRS time spectra recorded in remanence at different x-ray incidence angles $\alpha_{i}$ (data points: black symbols, fit: red curves). Labels state the azimuthal angles of the magnetization orientations in the two layer model.}
\label{fig:FigureSM10}
\end{minipage}
\end{figure}

Further NRS time spectra were recorded at both the R-plane and S-plane crystal truncation rod in remanence at $\alpha_{i} = 0.200^{\circ}$ and $\alpha_{i} = 0.275^{\circ}$ to probe deeper into the Fe film (see Fig.~\ref{fig:FigureSM10}). Considering that only a basic two layer model was assumed for fitting these spectra, a reasonable agreement between the fits and the data was achieved. The fitting parameters show qualitatively that the azimuthal magnetization orientation in the lower layer is closer to the easy axis orientation parallel to the facets ($\phi = 0$) than in the upper layer. This indicates a spring-like magnetization structure of the Fe film, where the top layers of the Fe film are coupled to the Cr capping layer, while the bottom layers are free to relax toward the easy axis orientation when no external field is applied.

